\documentclass[a4paper,11pt]{article}
\pdfoutput=1
\usepackage{jheppub}
\makeatletter
\renewcommand{\@fpheader}{}
\makeatother
\AddToHookNext{shipout/before}{\setbox\ShipoutBox=\vbox{\kern6mm\box\ShipoutBox}}
\usepackage{amsthm}
\usepackage{bm}
\usepackage{booktabs}

\newtheorem{proposition}{Proposition}
\newcommand{\dd}{\mathrm{d}}
\newcommand{\ii}{\mathrm{i}}
\newcommand{\Hint}{\hat{H}_{\rm int}}
\newcommand{\Hlab}{H_{\rm lab}}
\newcommand{\HFW}{H_{\rm FW}}
\newcommand{\HN}{H_{\rm N}}
\newcommand{\Htid}{H_{\rm tid}}
\newcommand{\cM}{\mathcal{M}}
\newcommand{\cA}{\mathcal{A}}
\newcommand{\cN}{\mathcal{N}}
\newcommand{\cH}{\mathcal{H}}

\newcommand{\cV}{\mathcal{V}}
\newcommand{\Mr}{\hat{M}_{\rm r}}
\newcommand{\Mi}{\hat{M}_{\rm i}}
\newcommand{\Mg}{\hat{M}_{\rm g}}
\newcommand{\bx}{\bm{x}}
\newcommand{\bX}{\bm{X}}
\newcommand{\bP}{\bm{P}}
\newcommand{\bK}{\bm{K}}
\newcommand{\bJ}{\bm{J}}
\newcommand{\bS}{\bm{S}}
\newcommand{\bC}{\bm{C}}
\newcommand{\ba}{\bm{a}}
\newcommand{\bom}{\bm{\varpi}}
\newcommand{\rinv}{r^{\rm inv}}

\title{Soft graviton theorems and the quantum equivalence principle}

\author[a,b]{Hiroki Matsui}

\affiliation[a]{Osaka Central Advanced Mathematical Institute (OCAMI), Osaka Metropolitan University,\\
3-3-138 Sugimoto, Sumiyoshi-ku, Osaka 558-8585, Japan}
\affiliation[b]{TRIP Headquarters, RIKEN,\\
2-1 Hirosawa, Wako, Saitama 351-0198, Japan}

\emailAdd{hiroki.matsui@yukawa.kyoto-u.ac.jp}

\abstract{The quantum equivalence principle states that the rest, inertial and gravitational mass operators of a system with internal energy levels are equal. We investigate how much of this statement follows from the on-shell consistency of a massless spin-two particle, treating stable internal levels as massive one-particle states at tree level. The leading and subleading soft theorems fix, respectively, the gravitational charge and the coupling to the Lorentz generators on each level and each degenerate multiplet, but do not determine the transitions between levels of different mass. For these transitions, we match the long-range exchange fixed by mass-changing on-shell amplitudes to a test Hamiltonian, assuming a regular local derivative expansion truncated at dipole order. The invariant off-diagonal coupling is then a combination of the potential and momentum-dependent terms, and it vanishes for gravity. Independently of the soft theorems, coupling to a local conserved stress tensor whose charges generate Poincar\'e transformations gives equal mass operators in a uniform weak field.}

\begin{document}
\maketitle
\flushbottom

\section{Introduction}
\label{sec:intro}

The Einstein equivalence principle (EEP) states that gravity couples universally to all forms of mass and energy~\cite{Will:2014kxa}. The inertial and gravitational masses of a body coincide, and every contribution to its internal energy, including the binding energy, enters both of them in the same way. For a system whose internal degrees of freedom are quantized, the internal energy is an operator $\Hint$. Zych and Brukner pointed out that the EEP then becomes an equality between operators~\cite{Zych:2015fka}. The rest, inertial and gravitational mass-energy operators,
\begin{equation}
\hat{M}_\alpha=m_\alpha\,\mathbf{1}+\frac{\hat{H}_{{\rm int},\alpha}}{c^2},\qquad \alpha={\rm r},\,{\rm i},\,{\rm g},
\label{eq:Malpha}
\end{equation}
must coincide, $\Mr=\Mi=\Mg$. We call this statement the quantum equivalence principle (QEP). We use hats for mass and internal-energy operators to distinguish them from their eigenvalues. Other operators, such as the Poincar\'e generators, are written without hats. Classical tests of the EEP use eigenstates of the internal energy and constrain only the diagonal elements of these operators. The off-diagonal elements, which would make $\Mg$ fail to commute with $\Hint$, have no classical analogue. The first experiment designed to constrain them compared the free fall of $^{87}$Rb atoms in two hyperfine states and in their coherent superposition, and reported the bound $|r|\lesssim5\times10^{-8}$ on the off-diagonal element of $\Mg\Mi^{-1}$~\cite{Rosi:2017ieh}. Further tests and related formulations of the equivalence principle for quantum systems have been discussed in refs.~\cite{Orlando:2015jdi,Das:2023cfu,Onofrio:2025ixt}. In the test theory used to interpret these experiments, the entries of $\Mg$ are free parameters.

On the other hand, from the viewpoint of the $S$-matrix, the classical equivalence principle is not an independent assumption. Weinberg showed that the Lorentz invariance of amplitudes with a soft graviton forces the graviton to couple to all particles with the same strength~\cite{Weinberg:1964ew,Weinberg:1965nx}. It is natural to ask whether the same consistency condition also fixes the operator form of the equivalence principle. Weinberg's argument relies on the pole of the propagator adjacent to the external line that emits the soft graviton. If the internal state of the system changes in the emission, from a level of mass $M_{n'}$ to a level of mass $M_n\neq M_{n'}$, this propagator has no pole at zero frequency. Hence, the leading soft theorem does not constrain the off-diagonal elements between levels of different energy. We therefore ask which representation-independent couplings are constrained by the soft theorems and by the factorization residues of mass-changing amplitudes, and what additional assumptions are required to match those constraints to a laboratory Hamiltonian.

The soft graviton factor is fixed by the four-momenta of the external particles at leading order~\cite{Weinberg:1964ew,Weinberg:1965nx} and, at tree level, by their total angular momenta at subleading order~\cite{Cachazo:2014fwa,Bern:2014vva,Sen:2017xjn,Sen:2017nim,Laddha:2017ygw}. For a particle that remains in the same internal level, the coupling to its spin is therefore universal, with no independent gravitational dipole moment~\cite{Falkowski:2020aso,Kobzarev:1962wt,Pagels:1966zza,Teryaev:1999su,Teryaev:2016edw,Silenko:2006er,Cotogno:2019xcl,Lorce:2019sbq}. Independent curvature couplings lie beyond the monopole and dipole orders considered here; their coefficients are not fixed by the leading and subleading theorems. For transitions between levels of different mass, the allowed on-shell three-point amplitudes with one massless particle have been classified~\cite{Conde:2016vxs,Arkani-Hamed:2017jhn}, and the corresponding stress-tensor matrix elements have been parametrized for specific spins~\cite{Kim:2022bwn,Ozdem:2022zig}.

The distinction between diagonal and transition matrix elements also arises for a bound system in a weak gravitational field. In the Newtonian gauge, the coefficient of the gravitational potential contains a virial term~\cite{Nordtvedt:1994gp,Carlip:1997gf}. Its expectation value vanishes in a stationary state, but its matrix elements between different energy levels need not vanish, which led to the claim that the gravitational mass operator of hydrogen does not commute with its internal Hamiltonian~\cite{Lebed:2011np,Lebed:2014oua}. When the internal dynamics is expressed in local variables, however, the gravitational mass is given by the total rest-frame energy divided by $c^2$~\cite{Zych:2018bmk}, and explicit weak-field Hamiltonians of atoms contain the same mass in the kinetic and potential terms\nocite{Schwartz:2019qqg,Hartong:2023yxo}~\cite{Schwartz:2019qqg,Janson:2024hdt}. These Hamiltonians are derived for particular binding models and coordinate systems. This motivates us to distinguish the coefficient that is called the gravitational mass in a given representation from the invariant transition data that can be matched to on-shell amplitudes.

Our analysis has two logically distinct parts. The first uses on-shell amplitudes to constrain universal charges and the long-range part of single-graviton exchange. Its connection to a low-energy test Hamiltonian is made through the invariant transition coefficient in section~\ref{sec:invariant}, with the matching assumptions stated in section~\ref{sec:assumptions}. The second assumes a local conserved stress tensor whose charges are the Poincar\'e generators and independently derives the laboratory Hamiltonian that contains the same low-gradient couplings. This second derivation does not rely on the soft theorems. Our results are summarized as follows.
\begin{itemize}
\item[(i)] The soft pole fixes the gravitational charge matrix to be proportional to the identity on every set of degenerate states that scatter with each other, but gives no information about transitions between levels of different mass. The transition vertex at zero momentum transfer is not an observable of the $S$-matrix, because it can be traded for a contact term (section~\ref{sec:soft}).
\item[(ii)] The on-shell transition amplitude is the residue of the pole at which the other level is on its mass shell. Lorentz invariance excludes a transition between spinless levels, gives the necessary selection rule $s_n+s_{n'}\ge2$ for the spins of the two levels and permits a curvature representation of every allowed transition residue. Factorization then fixes the corresponding massless exchange pole (section~\ref{sec:pole}). Translating this representation into a tidal potential requires the regularity and derivative-expansion assumptions stated in section~\ref{sec:tchannel}.
\item[(iii)] At first order in the soft momentum, gauge invariance expresses the sum of the regular terms through the amplitude of the same levels and fixes the vertex on a degenerate multiplet to be the Lorentz generator of the one-particle states. This excludes an anomalous gravitomagnetic moment, a gravitoelectric dipole coupling of the spin and dipole transitions within a degenerate multiplet. These are constraints on the on-shell monopole and dipole data. Their relation to the value and gradient of a static potential requires matching (section~\ref{sec:sub}).
\item[(iv)] If the graviton couples to a local conserved source, the complete one-graviton vertex between spinless levels of different mass is a coupling to the linearized Ricci tensor, which vanishes for an on-shell graviton and gives only contact terms with a static source in the local derivative expansion (section~\ref{sec:source}).
\item[(v)] Under the additional local-source assumption, in the proper reference frame of a uniformly accelerated and rotating laboratory, the Hamiltonian is $H+\boldsymbol a\cdot\boldsymbol K-\boldsymbol\varpi\cdot\boldsymbol J$, where $H$, $\boldsymbol K$ and $\boldsymbol J$ are the Poincar\'e generators of the isolated system. It commutes with the invariant-mass operator $\hat M$. On the space of stable levels the same operator $\hat M$ appears in the rest energy, the kinetic term and the inertial potential, and the internal structure enters only at tidal order (section~\ref{sec:matching}).
\item[(vi)] The virial terms in the Newtonian gauge are removed by a unitary transformation generated by the dilatation and special conformal moments of the stress tensor, for any matter theory with a conserved symmetric stress tensor and any value of the post-Newtonian parameter $\gamma$. The off-diagonal elements of $\hat{M}_{\rm g}$ between non-degenerate levels are not invariant under field-dependent changes of the internal basis. Within the Hamiltonian family of eq.~\eqref{eq:family} and its retained order, the on-shell transition amplitude depends on the invariant combination~\eqref{eq:rinv}. Matching the non-tidal gravitational sector gives $\rinv_{mn}=0$. The local-source construction independently reproduces this result, together with the universal redshift of the internal spectrum and the vanishing of the covariant non-adiabatic coupling in a uniform field (section~\ref{sec:rep}).
\item[(vii)] Additional light fields with charges that are not Poincar\'e generators can produce long-range violations below tidal order (section~\ref{sec:fields}). In a two-level model with a static off-diagonal coupling, rapid internal evolution suppresses the first-order interferometric response, and level repulsion produces a level-dependent force at second order (section~\ref{sec:obs}).
\end{itemize}

The central connection to the test theory is the matching in section~\ref{sec:invariant}. In the specified family of Hamiltonians, an off-diagonal entry of $\Mg$ alone is not invariant: a field-dependent change of internal basis also generates a momentum-dependent interaction. The combination of these coefficients that enters the on-shell transition amplitude is invariant, and its vanishing is the physical constraint in the non-tidal sector. Throughout the paper, we distinguish the pole data, the assumptions needed for their derivative expansion and the independent local-source derivation of the laboratory Hamiltonian.

The ingredients of (i)--(iii), including universal diagonal couplings, the absence of independent gravitational dipole moments and the three-point selection rule, are established results. We apply them to the internal levels of a single system and follow their implications through the matching. The normalization of stress-tensor matrix elements by Poincar\'e charges is also known for arbitrary spin~\cite{Cotogno:2019xcl,Lorce:2019sbq}, and mass-energy equivalence in appropriate local variables has been studied for composite systems~\cite{Zych:2018bmk,Schwartz:2019qqg,Janson:2024hdt,Pikovski:2013qwa}. Our local-source construction gives the laboratory Hamiltonian explicitly and clarifies its relation to the Newtonian gauge and to the invariant data of the test theory. The local-source hypothesis itself is not derived from the soft theorems.

The amplitude analysis is at tree level, with stable external levels and fixed non-zero mass gaps in the non-degenerate sector. In four dimensions, massless loops give logarithmic terms in the soft expansion and non-local corrections at finite momentum transfer~\cite{Sahoo:2018lxl,Donoghue:1994dn,Bjerrum-Bohr:2014zsa}. The matching further assumes a controlled local derivative expansion and the stated operator truncation. The independent source derivation holds at first order in the external field, below tidal order. The second-order effects in section~\ref{sec:obs} are predictions of the two-level Hamiltonian specified there.

We use the metric signature $(-,+,+,+)$. In sections~\ref{sec:soft}--\ref{sec:source} we set $\hbar=c=1$. From section~\ref{sec:matching} on we restore $\hbar$ and $c$. Latin indices $i,j,k$ denote spatial components and repeated indices are summed. In sections~\ref{sec:soft}--\ref{sec:sub} the index $i$ also labels the external particles of an amplitude.

\section{Setup and assumptions}
\label{sec:setup}

\subsection{Test theory of the quantum equivalence principle}

We first recall the test theory of refs.~\cite{Zych:2015fka,Rosi:2017ieh}. A system with internal degrees of freedom in a weak gravitational potential $\Phi$ is described, at low velocities, by the Hamiltonian
\begin{equation}
H_{\rm test}=\Mr c^2+\frac{\bP^2}{2\Mi}+\Mg\,\Phi(\bX),
\label{eq:Htest}
\end{equation}
where $\bX$ and $\bP$ are the position and momentum of the center of mass and the three operators $\hat{M}_\alpha$ of eq.~\eqref{eq:Malpha} act on the internal Hilbert space $\cH_{\rm int}$. The equality $\Mr=\Mi$ corresponds to local Lorentz invariance, $\Mr=\Mg$ to local position invariance and $\Mi=\Mg$ to the weak equivalence principle~\cite{Zych:2015fka}. If the three operators have the same spectrum but do not commute, the violation has no classical counterpart. For two internal levels $|1\rangle$ and $|2\rangle$ the weak equivalence principle is parametrized as~\cite{Rosi:2017ieh}
\begin{equation}
\Mg\Mi^{-1}\simeq\begin{pmatrix} r_1 & r\\ r^* & r_2\end{pmatrix},
\label{eq:rmatrix}
\end{equation}
where the product is treated as a Hermitian matrix at the lowest order in $1/c^2$. The classical weak equivalence principle requires $r_1=r_2=1$, and the quantum one also requires $r=0$. The experiment of ref.~\cite{Rosi:2017ieh} compared the free fall of the hyperfine states $|1\rangle=|F=1,m_F=0\rangle$ and $|2\rangle=|F=2,m_F=0\rangle$ of $^{87}$Rb, whose transition frequency is $\Omega/2\pi\simeq6.83\,{\rm GHz}$, and of their coherent superposition. The bound $|r|\lesssim5\times10^{-8}$ was obtained from the measurements with the superposition.

\subsection{Stable levels as one-particle states}

We consider a Poincar\'e-invariant quantum theory. A stable level of a bound system, such as a hyperfine level of an atom in its electronic ground state, is a one-particle state. We denote these states by $|n,\bm{p},\sigma\rangle$, where $n$ labels the level, $\bm{p}$ is the momentum and $\sigma$ is the spin component. For each $n$ they span an irreducible unitary representation of the Poincar\'e group with mass $M_n$ and spin $s_n$~\cite{Weinberg:1995mt}. The generators of time translations, space translations, rotations and boosts, $H$, $\bP$, $\bJ$ and $\bK$, act within each irreducible representation. We define the mass operator by
\begin{equation}
\hat{M}^2=\frac{H^2}{c^4}-\frac{\bP^2}{c^2},\qquad \hat{M}=\sum_n M_n\,\Pi_n,\qquad M_n=m+\frac{E_n}{c^2},
\label{eq:Mop}
\end{equation}
where $\Pi_n$ is the projector onto the level $n$, $E_n$ is its excitation energy and $m$ is the mass of the lowest level. The internal Hilbert space $\cH_{\rm int}$ is spanned by states labeled by $(n,\sigma)$, and the internal Hamiltonian is $\Hint=\sum_nE_n\Pi_n$, so that $\hat{M}=m\mathbf{1}+\Hint/c^2$. From the viewpoint of a graviton whose wavelength is much larger than the size of the system, each level is an elementary particle. This is the starting point of the soft graviton effective theory of ref.~\cite{Sundrum:2003jq}.

In each level the energy is $H=(M_n^2c^4+c^2\bP^2)^{1/2}=M_nc^2+\bP^2/2M_n+\cdots$. Hence, the operator that generates the internal evolution in the rest frame and the operator that appears in the kinetic term are both equal to $\hat{M}$. The equality $\Mr=\Mi$ follows from Poincar\'e invariance of the one-particle states and does not involve gravity. The non-trivial part of the QEP is the identification of $\Mg$.

\subsection{Assumptions}
\label{sec:assumptions}

We distinguish three sets of assumptions.

\paragraph{On-shell amplitudes.}
We consider stable one-particle levels in a unitary, Poincar\'e-invariant theory with a massless graviton of helicity $\pm2$. We treat metastable levels as stable when their lifetimes are much longer than the duration of the experiment. In sections~\ref{sec:soft}--\ref{sec:sub} we work at the lowest order in the graviton coupling $\kappa$. We assume that the amplitudes for single-graviton emission or exchange have one-particle poles whose residues factorize into on-shell amplitudes, and that they are otherwise analytic in the region of the expansion. We isolate the massless spin-two exchange; additional light mediators are discussed in section~\ref{sec:fields}. Non-zero internal mass gaps are held fixed in the soft limit. In four dimensions the leading theorem also holds for infrared-regulated amplitudes~\cite{Weinberg:1965nx}, while loops modify the subleading expansion (section~\ref{sec:loops}).

\paragraph{Matching to a test Hamiltonian.}
To interpret the pole data as couplings to a static potential, we assume a local, long-wavelength and low-velocity expansion whose coefficients are regular in the momentum transfer at each fixed non-zero mass gap. Curvature operators and terms supported on the source are kept as separate sectors before the matter equations of motion are used. The monopole--dipole test family in section~\ref{sec:invariant} is a truncation that is specified after the universal orbital and spin couplings have been separated. Its use requires that the omitted momentum-dependent or tensor operators do not contribute at the retained order. A curvature representation of a three-point residue alone does not establish these closure and power-counting assumptions for arbitrary spin. We do not consider a limit in which the soft momentum and the mass gap are taken to zero simultaneously.

\paragraph{Local-source construction.}
From section~\ref{sec:source} onward, where explicitly stated, we additionally assume that the graviton couples to a local, symmetric and conserved stress tensor $T^{\mu\nu}$ of the matter system. Its charges generate Poincar\'e transformations, and its coupling to the metric gives the linear response in eq.~\eqref{eq:Hh}. This identification of the source is an additional physical input and is not a consequence of the soft theorems. Together with Poincar\'e symmetry, it independently determines the laboratory response at first order in the field when curvature across the system is neglected.

\section{Constraints from the soft pole}
\label{sec:soft}

\subsection{Leading soft factor with internal labels}

Let $\cM_N$ be an amplitude with $N$ external particles. We take all momenta $k_i$ as outgoing, so that an incoming particle of physical momentum $p_i$ has $k_i=-p_i$, and $k_i^2=-M_{n_i}^2$ for a particle in the level $n_i$. Each external particle may be a level of a composite system, and $\cM_N$ carries the corresponding internal labels. We consider the same process with an additional outgoing graviton of momentum $q=\omega(1,\hat{\bm{q}})$, where $\hat{\bm{q}}$ is the unit propagation vector. Its polarization $\varepsilon_{\mu\nu}$ is symmetric, traceless and transverse, $\varepsilon_{\mu\nu}q^\nu=0$. The singular terms for $\omega\to0$ arise when the graviton is attached to an external line. If the internal state of the particle $i$ changes from the level $n'$ to the level $n$ in the emission, the adjacent propagator is
\begin{equation}
\frac{1}{(k_i+q)^2+M_{n'}^2}=\frac{1}{2k_i\cdot q+\Delta_{n'n}},\qquad \Delta_{n'n}\equiv M_{n'}^2-M_n^2 .
\label{eq:propagator}
\end{equation}
It has a pole at $\omega=0$ only if $M_{n'}=M_n$. For $M_{n'}\neq M_n$ the contribution is regular at $\omega=0$. Hence, the leading soft factor is block diagonal in the mass.

Within a degenerate block the residue of the pole is the vertex between two states of the same mass at vanishing momentum transfer. By Lorentz covariance it is proportional to $\varepsilon_{\mu\nu}k_i^\mu k_i^\nu$, since any dependence on the spin requires at least one power of $q$. The proportionality factor is a matrix $G$ acting on the degenerate labels, which we call the gravitational charge matrix. It commutes with the spin, and it is Hermitian by the unitarity of the $S$-matrix. The leading soft theorem reads
\begin{equation}
\cM_{N+1}=\frac{\kappa}{2}\sum_i\frac{\varepsilon_{\mu\nu}k_i^\mu k_i^\nu}{k_i\cdot q}\,G^{(i)}\cM_N+O(\omega^0),
\label{eq:softG}
\end{equation}
where $\kappa^2=32\pi G_{\rm N}$ and $G^{(i)}$ acts on the internal label of the particle $i$. We do not assume $G=\mathbf{1}$. The index $i$ in the sum labels the external line to which the graviton is attached. On that line, the matrix element $G^{(i)}_{a_i b_i}$ connects the internal state $b_i$ in the hard amplitude $\cM_N$ to the state $a_i$ in $\cM_{N+1}$. The sum over $b_i$ is implicit in the product $G^{(i)}\cM_N$, with all other internal labels held fixed.

The amplitude of a massless particle of helicity $\pm2$ is Lorentz invariant only if it is invariant under $\varepsilon_{\mu\nu}\to\varepsilon_{\mu\nu}+q_\mu\xi_\nu+q_\nu\xi_\mu$ with $\xi\cdot q=0$~\cite{Weinberg:1964ew}. Applying this shift to eq.~\eqref{eq:softG} gives the condition
\begin{equation}
\sum_i(\xi\cdot k_i)\,G^{(i)}\cM_N=0
\label{eq:gaugecond}
\end{equation}
for all such $\xi$ and all kinematics. Since $G$ is Hermitian, we can choose the external states to be eigenstates of $G$ with eigenvalues $g_{a_i}$. The condition becomes $\sum_ig_{a_i}k_i^\mu=0$ for every process with $\cM_N\neq0$. For generic momenta the only linear relation among the $k_i$ is the conservation law $\sum_ik_i^\mu=0$. Therefore, all the eigenvalues that appear in a non-vanishing amplitude are equal. We conclude that
\begin{equation}
G=g\,\mathbf{1}
\label{eq:Guniversal}
\end{equation}
on every set of states that are connected by scattering, with a common constant $g$ that is absorbed into the definition of $\kappa$. States with different eigenvalues of $G$ would not interact with each other.

Equation~\eqref{eq:Guniversal} contains two statements. The diagonal elements are equal for all levels, which is Weinberg's theorem applied to each level as a separate particle. In addition, the gravitational charge has no off-diagonal elements between degenerate states. This is a straightforward extension of the original argument, of the same kind as the one that leads to matrix-valued charges for soft gluons. The leading soft factor is not renormalized~\cite{Weinberg:1965nx,Bern:2014oka}, and the argument refers only to the asymptotic states. Hence, eq.~\eqref{eq:Guniversal} does not depend on the dynamics that binds the system. In the rest frame of a level the soft factor is $-\kappa\,\varepsilon_{00}M_n/2\omega$. As an operator on $\cH_{\rm int}$ the charge to which the soft graviton couples is therefore the mass operator $\hat{M}$ of eq.~\eqref{eq:Mop}.

\subsection{Transitions between non-degenerate levels}
\label{sec:softnondeg}

To discuss the terms that are regular at $\omega=0$, we write the amplitude as $\cM_{N+1}=\tfrac{\kappa}{2}\varepsilon_{\mu\nu}\cM^{\mu\nu}$ and decompose it as
\begin{align}
\cM^{\mu\nu}={}&\sum_i\frac{\cV_i^{\mu\nu}(k_i,q)}{k_i\cdot q}\,\cM_N(\ldots,k_i+q,\ldots)\nonumber\\
&+\sum_i\sum_{n'\neq n_i}\frac{2V^{\mu\nu}_{n_in'}(k_i,q)}{2k_i\cdot q+\Delta_{n'n_i}}\,\cM_N^{(i:n')}(\ldots,k_i+q,\ldots)+N^{\mu\nu}(q).
\label{eq:decomp}
\end{align}
The first term is the emission from an external line without a change in mass, with $\cV_i^{\mu\nu}=k_i^\mu k_i^\nu+O(q)$ by eq.~\eqref{eq:Guniversal}. In the second term the particle $i$ is produced in the level $n'$ and makes a transition to the level $n_i$ by emitting the graviton. Here $V^{\mu\nu}_{nn'}$ is the transition vertex, $\cM_N^{(i:n')}$ denotes the amplitude in which the particle $i$ is in the level $n'$, and a sum over the spin states of the intermediate level is understood. The last term $N^{\mu\nu}$ is the emission from the internal lines, which is analytic at $q=0$ in the tree-level expansion.

For $\omega\to0$ the second term of eq.~\eqref{eq:decomp} gives a contribution of order $\omega^0$,
\begin{equation}
\frac{\kappa}{2}\,\varepsilon_{\mu\nu}\,\frac{2V^{\mu\nu}_{n_in'}(k_i,0)}{\Delta_{n'n_i}}\,\cM_N^{(i:n')} ,
\label{eq:mixingterm}
\end{equation}
which is proportional to the amplitude for the production of another level. It has the appearance of a mixing of the levels induced by gravity. However, this term has no invariant meaning. In the soft limit the intermediate state is off its mass shell by the fixed amount $\Delta_{n'n_i}$, and its propagator cannot be distinguished from a contact interaction. As an example, let two spinless levels be described by scalar fields $\phi_n$ and $\phi_{n'}$ that are minimally coupled to the metric $g_{\mu\nu}=\eta_{\mu\nu}+\kappa h_{\mu\nu}$, so that $V^{\mu\nu}_{nn'}=0$. The redefinition
\begin{equation}
\phi_{n'}\;\to\;\phi_{n'}+\frac{b\kappa}{2}\,h^{\mu\nu}\partial_\mu\partial_\nu\phi_n
\label{eq:redefinition}
\end{equation}
does not change the $S$-matrix~\cite{Chisholm:1961tha,Kamefuchi:1961sb}. In the new variables the kinetic term of $\phi_{n'}$ gives a transition vertex that is proportional to the inverse propagator, and the interaction terms give contact terms,
\begin{equation}
V^{\mu\nu}_{nn'}(k,q)=\frac{b}{2}\,k^\mu k^\nu\big(2k\cdot q+\Delta_{n'n}\big),\qquad
N^{\mu\nu}\supset-b\sum_ik_i^\mu k_i^\nu\,\cM_N^{(i:n')}(\ldots,k_i+q,\ldots).
\label{eq:redefvertex}
\end{equation}
Each of the two contributions to eq.~\eqref{eq:decomp} is of order $\omega^0$ and is proportional to the amplitude of the other level, and their sum vanishes. We give the details for an explicit model in appendix~\ref{app:toy}. Conversely, since $\Delta_{n'n}\neq0$, any transition vertex at $q=0$ can be written as a multiple of the inverse propagator and can be moved into $N^{\mu\nu}$.

We conclude that the transition vertex at vanishing momentum transfer between levels of different mass is not an observable of the $S$-matrix. For this reason the absence of a pole in eq.~\eqref{eq:propagator} cannot be used to conclude that gravity does not mix internal levels, and the term~\eqref{eq:mixingterm} cannot be used to conclude that it does. We will see the static counterpart of this fact in section~\ref{sec:rep}, where the Hamiltonian in the Newtonian gauge has off-diagonal elements that are absent in the proper reference frame. The invariant information about the transitions is contained in two places. The first is the residue of the pole of the second term of eq.~\eqref{eq:decomp} at $2k_i\cdot q=-\Delta_{n'n_i}$, which we study in section~\ref{sec:pole}. The second is the sum of the regular terms, which is constrained by gauge invariance and which we study in section~\ref{sec:sub}.

\section{The pole on the mass shell of another level}
\label{sec:pole}

\subsection{Factorization and the on-shell transition amplitude}

For $M_{n'}>M_n$ the second term of eq.~\eqref{eq:decomp} has a pole at $(k_i+q)^2=-M_{n'}^2$, where the particle $i$ is produced in the level $n'$ and then emits the graviton. In the rest frame of the intermediate state the energy of the graviton at the pole is
\begin{equation}
\omega_*=\frac{M_{n'}^2-M_n^2}{2M_{n'}},
\label{eq:omegastar}
\end{equation}
which is the frequency of the transition. By unitarity the amplitude factorizes at the pole,
\begin{equation}
\cM_{N+1}\;\to\;\sum_{\sigma'}\frac{A_3(n',\sigma'\to n+h)\;\cM_N^{(i:n',\sigma')}}{2k_i\cdot q+\Delta_{n'n}},\qquad
A_3=\kappa\,\varepsilon_{\mu\nu}V^{\mu\nu}_{nn'}(k_i,q)\Big|_{2k_i\cdot q=-\Delta_{n'n}},
\label{eq:factorization}
\end{equation}
where $A_3$ is the amplitude for the decay of the level $n'$ into the level $n$ and a graviton. The width of the level $n'$ due to this decay is of order $\kappa^2$ and is neglected at the order at which we work. For $M_{n'}<M_n$ the same amplitude describes the absorption of a graviton. In both cases $A_3$ is the amplitude of a physical process with real momenta. It must therefore be Lorentz invariant by itself, that is, invariant under $\varepsilon_{\mu\nu}\to\varepsilon_{\mu\nu}+q_\mu\xi_\nu+q_\nu\xi_\mu$. This is Weinberg's condition, applied at the pole~\eqref{eq:factorization} instead of the pole at $\omega=0$.

\subsection{Selection rule and Riemann form}
\label{sec:selection}

Let us first consider two spinless levels. The only available vectors are $k$ and $q$. Since $\varepsilon_{\mu\nu}q^\nu=0$ and $\varepsilon^\mu{}_\mu=0$, the most general amplitude is
\begin{equation}
A_3=a\,\kappa\,\varepsilon_{\mu\nu}k^\mu k^\nu
\label{eq:A3scalar}
\end{equation}
with a constant $a$, which would be a transition monopole. Under the polarization shift, the amplitude changes by
\begin{equation}
\delta A_3=2a\kappa\,(k\cdot q)(\xi\cdot k)=-a\kappa\,\Delta_{n'n}\,(\xi\cdot k).
\label{eq:deltaA3}
\end{equation}
For $\Delta_{n'n}=0$ this vanishes, and $a$ is the charge $g$ of section~\ref{sec:soft}. For $\Delta_{n'n}\neq0$ it does not vanish, and Lorentz invariance requires $a=0$. Hence, a transition monopole between two levels of different mass is not a Lorentz-invariant coupling of a massless particle of helicity two, and the pole~\eqref{eq:factorization} is absent for spinless levels. Together with eq.~\eqref{eq:Guniversal}, this excludes an off-diagonal monopole coupling for all values of $\Delta_{n'n}$. The statement at $\Delta_{n'n}\neq0$ concerns the residue of a pole and does not rely on the expansion in $\omega$, which is valid only for $\omega\ll\omega_*$.

For levels with spin, conservation of angular momentum gives a necessary selection rule. In the rest frame of $n'$ the graviton is emitted with helicity $\lambda_h=\pm2$ along some direction and the level $n$ recoils in the opposite direction with helicity $\lambda_n$. The component of the angular momentum along the direction of emission is $\lambda_h-\lambda_n$~\cite{Jacob:1959at}, and a state of $n'$ with this component exists only if $|\lambda_h-\lambda_n|\le s_{n'}$. With $|\lambda_n|\le s_n$ this is possible only if
\begin{equation}
s_{n'}+s_n\ge2 .
\label{eq:selection}
\end{equation}
In particular, two levels with spins $0$ and $0$, $\tfrac12$ and $\tfrac12$, or $0$ and $1$ are not connected by the emission or absorption of a single graviton. The condition~\eqref{eq:selection} is the special case $|h|=2$ of the constraint $s_1+s_2\ge|h|$ on the three-point amplitudes of two particles of different mass and a massless particle of helicity $h$~\cite{Conde:2016vxs,Arkani-Hamed:2017jhn}. These amplitudes are the building blocks that are used to describe the absorption of gravitational waves by black holes in ref.~\cite{Aoude:2023fdm}, where they are called mass-changing amplitudes.

When the condition~\eqref{eq:selection} is satisfied, an allowed amplitude admits a representation in terms of the linearized curvature on the transition mass shell. To see this, we use the following algebraic identity. Let $q$ be non-zero and let $\ell$ be a vector that satisfies $\ell\cdot q=1$. Then a symmetric polarization can be written as
\begin{equation}
\varepsilon_{\mu\sigma}=-2\ell^\nu\ell^\rho R_{\mu\nu\rho\sigma}[\varepsilon]+q_\mu\zeta_\sigma+q_\sigma\zeta_\mu,\qquad
\zeta_\sigma=\ell^\rho\varepsilon_{\rho\sigma}-\frac12q_\sigma\,\ell^\rho\ell^\lambda\varepsilon_{\rho\lambda},
\label{eq:lemma}
\end{equation}
where $R_{\mu\nu\rho\sigma}[\varepsilon]$ is the linearized Riemann tensor of the plane wave $h_{\mu\nu}=\varepsilon_{\mu\nu}e^{\ii q\cdot x}$, given in appendix~\ref{app:conventions}. At the pole we have $k\cdot q=-\Delta_{n'n}/2$ and can take $\ell=-2k/\Delta_{n'n}$. Since $A_3$ is invariant under the shift of the polarization, the last two terms of eq.~\eqref{eq:lemma} do not contribute, and
\begin{equation}
A_3=-\frac{8\kappa}{\Delta_{n'n}^2}\,V^{\mu\sigma}_{nn'}\,k^\nu k^\rho\,R_{\mu\nu\rho\sigma}[\varepsilon].
\label{eq:RiemannForm}
\end{equation}
In the rest frame of the level $n$ this is $(8\kappa M_n^2/\Delta_{n'n}^2)\,V^{jl}_{nn'}R_{0j0l}$, a coupling of a symmetric tensor operator on $\cH_{\rm int}$ to the electric components of the curvature, which is the form of a tidal interaction. Equation~\eqref{eq:RiemannForm} is an on-shell identity that follows from Lorentz invariance. It neither determines $V^{jl}_{nn'}$, which depends on the structure of the system, nor by itself establishes a regular local gradient expansion. The factor $1/\Delta_{n'n}^2$ is finite at a fixed non-zero gap, but the representation is not uniform in the degenerate limit. Moreover, for unequal masses the limit $q^\mu\to0$ cannot be taken with both matter states kept on shell. These points must be kept in mind when this identity is used for the static matching. For the minimal amplitudes of ref.~\cite{Aoude:2023fdm} the assumption of locality gives a coupling that vanishes as a power of the mass difference, in agreement with the fact that the corresponding coupling within a degenerate multiplet is of second order in $q$ (section~\ref{sec:sub}).

\subsection{From the exchange pole to the long-range interaction}
\label{sec:tchannel}

The object to be matched is a four-point exchange amplitude, not a soft factor evaluated at a static momentum. A radiated graviton has $q^2=0$, whereas a static exchanged graviton has $q^0=0$ and generally $q^2=\bm q^2\neq0$. Factorization fixes the residue at $q^2=0$, and a regular continuation together with a derivative expansion then provides a representative of the interaction away from the pole.

We consider the scattering of the system off a heavy source $S$ of mass $M_S$, in which the system changes from level $n$ to level $n'$. Let $k$ and $k'$ be the physical initial and final momenta of the system, and $P$ and $P'$ those of the source. The exchanged momentum is $q=k'-k=P-P'$, and $t=-q^2$. The two mass-shell conditions give
\begin{equation*}
2P\cdot q=q^2,\qquad 2k\cdot q=-\Delta_{n'n}-q^2.
\end{equation*}
In the source rest frame the first relation gives $q^0=-q^2/(2M_S)$. Neglecting the recoil of the source, we have $q^0=0$ and $t=-\bm q^2$. The system then has the same total energy before and after scattering, with the change in kinetic energy compensating the change in rest energy. For unequal masses, $q^\mu=0$ is impossible on shell. The pole at $t=0$ is reached by analytic continuation to a non-zero complex null momentum transfer, where $2k\cdot q=-\Delta_{n'n}$ as in section~\ref{sec:selection}.

At order $\kappa^2$, using the propagator convention of eq.~\eqref{eq:factorization}, the one-graviton exchange amplitude factorizes as
\begin{equation}
\cM_4=\frac{\mathcal{R}_{n'n}}{q^2}+\cM_{\rm reg},
\qquad
\mathcal{R}_{n'n}
=\left.\sum_{\lambda=\pm2}
 A_3(n+h_\lambda\to n')\,A_3(S\to S+h_\lambda)
\right|_{t=0}.
\label{eq:tresidue}
\end{equation}
The graviton is outgoing on the source side and incoming on the transition side. The residue in $t$ differs by an overall sign since $q^2=-t$. The remainder $\cM_{\rm reg}$ is analytic at $t=0$ under the assumptions of section~\ref{sec:assumptions}. Two regular continuations of the residue away from the pole can differ by a term proportional to $t$, which contributes only to the regular remainder. In the static low-momentum expansion, the Fourier transform of a polynomial in $\bm q$ gives derivatives of a delta function for a point source. For a finite source these terms are local in the source profile at each order in the expansion.

To obtain the field carried by the pole, we keep the leading static vertex of the source, which is proportional to $P_\alpha P_\beta$ by the result of section~\ref{sec:soft}. At $t=0$ this vertex is conserved because $P\cdot q=0$. The transition vertex can be represented by eq.~\eqref{eq:RiemannForm}. The linearized Riemann tensor annihilates a pure-gauge polarization, so that the physical helicity sum in eq.~\eqref{eq:tresidue} gives the same residue as the covariant spin-two numerator
\begin{equation*}
\Pi_{\mu\nu,\alpha\beta}=
\frac12\bigl(\eta_{\mu\alpha}\eta_{\nu\beta}
+\eta_{\mu\beta}\eta_{\nu\alpha}
-\eta_{\mu\nu}\eta_{\alpha\beta}\bigr).
\end{equation*}
For a static source with four-velocity $u^\mu=(1,\bm 0)$, the contraction is $\Pi_{\mu\nu,\alpha\beta}u^\alpha u^\beta=u_\mu u_\nu+\eta_{\mu\nu}/2$. Normalizing the pole by the Newtonian potential $\Phi$ gives the metric representative $h_{00}=-2\Phi$, $h_{ij}=-2\Phi\delta_{ij}$ and $h_{0i}=0$. In particular, the linearized curvature of appendix~\ref{app:conventions} gives
\begin{equation*}
R_{0i0j}=-\frac12\partial_i\partial_j h_{00}
=\partial_i\partial_j\Phi .
\end{equation*}
All non-zero components of the curvature of this static metric are linear in the second spatial derivatives of $\Phi$.

To proceed from the residue to a local interaction, we now use the matching assumptions of section~\ref{sec:assumptions}. At fixed $\Delta_{n'n}\neq0$, the explicit factor $1/\Delta_{n'n}^2$ introduces no inverse power of the momentum transfer, provided that the remaining coefficient admits the assumed regular continuation. Each explicit spatial momentum carried by the field then corresponds to a derivative of the field, before the on-shell relations of the matter states are used. With the Fourier convention $f(\bm x)=\int\dd^3\bm q\,e^{\ii\bm q\cdot\bm x}\tilde f(\bm q)/(2\pi)^3$, the leading static term in this representative has the form
\begin{equation*}
\widetilde U_{n'n}(\bm q)
=-C^{ij}_{n'n}q_iq_j\tilde\Phi(\bm q),
\qquad
U_{n'n}(\bm x)=C^{ij}_{n'n}\partial_i\partial_j\Phi(\bm x).
\end{equation*}
Here $C^{ij}_{n'n}$ is the symmetric transition coefficient at the leading retained order in the velocity of the system, and it includes the normalization factor that relates the amplitude to a potential. More generally, the coefficient can depend on the momentum and spin of the system, and the corresponding operator requires Hermitian ordering. The above expression is not a classification of all such operators. The trace part gives $C^{ii}_{n'n}\nabla^2\Phi/3$. In momentum space its factor of $\bm q^2$ cancels the propagator. In position space it is supported on the source because $\nabla^2\Phi=4\pi G_{\rm N}\rho_S$, where $\rho_S$ is the source mass density. Outside the source only the traceless part of $C^{ij}_{n'n}$ contributes. For a point mass, the potential in momentum space and its Hessian outside the source are
\begin{equation*}
\tilde\Phi(\bm q)=-\frac{4\pi G_{\rm N} M_S}{\bm q^2},
\qquad
\partial_i\partial_j\Phi(\bm x)
=\frac{G_{\rm N} M_S}{r^3}
\bigl(\delta_{ij}-3\hat x_i\hat x_j\bigr),
\qquad r=|\bm x|>0 .
\end{equation*}
The potential $U_{n'n}$ is therefore of tidal type and falls off as $1/r^3$ outside the source. Its coefficient can vanish, as it does for $s_n+s_{n'}<2$ by eq.~\eqref{eq:selection}. Within the regular derivative expansion, the transition pole thus admits a representative with no independent coupling to $\Phi$ or $\nabla\Phi$. This is a statement about a representative and its power counting, not about every coefficient in a Hamiltonian obtained after field redefinitions or the use of on-shell identities.

The derivative order cannot be identified directly on the mass shell of the unequal-mass states. At $t=0$ we have $k\cdot q=-\Delta_{n'n}/2$, so that contractions containing $q$ can be replaced by the fixed mass gap. As a result, a momentum-dependent curvature operator can look like a coupling with fewer gradients after the on-shell conditions are used. The fixed-gap identity~\eqref{eq:RiemannForm} alone does not justify discarding such an operator. The invariant matching in section~\ref{sec:invariant} therefore applies to an explicitly closed monopole--dipole truncation, in which the curvature contributions are separated before the on-shell conditions are used. It is not a general classification for arbitrary spin or for a simultaneous zero-gap limit.

For two spinless or two spin-$\tfrac12$ levels, a stronger statement follows directly from eq.~\eqref{eq:selection}: the transition three-point amplitude, and hence its one-graviton exchange residue, vanishes. This statement about the pole does not require a choice of potential. It constrains, for example, proposals for an off-diagonal gravitational monopole of neutrinos~\cite{Gasperini:1988zf} under our on-shell assumptions. For transitions allowed by the spin selection rule, the curvature sector must be kept separate. Terms that are analytic at the exchange pole are local in the source expansion, whereas massless loops can add further non-analytic long-range terms~\cite{Donoghue:1994dn,Bjerrum-Bohr:2014zsa}. The additional source hypothesis in section~\ref{sec:source} gives information about the local terms as well.

\section{Constraints at subleading order}
\label{sec:sub}

\subsection{Gauge invariance at first order in the soft momentum}
\label{sec:low}

We now expand eq.~\eqref{eq:decomp} to first order in $q$. On a degenerate block the vertex can be written as
\begin{equation}
\cV_i^{\mu\nu}(k,q)=k^\mu k^\nu+\frac12\big(k^\mu A_i^{\nu\rho}+k^\nu A_i^{\mu\rho}\big)q_\rho+O(q^2),\qquad A_i^{\nu\rho}=-A_i^{\rho\nu},
\label{eq:vertexq}
\end{equation}
where $A_i^{\nu\rho}$ is a matrix that may depend on $k$ and acts on the spin and degenerate labels of particle $i$. This is the general form at this order. Terms proportional to $q^\mu$, $q^\nu$ or $\eta^{\mu\nu}$ do not contribute for a transverse and traceless polarization. Terms proportional to $k\cdot q$ cancel the pole and belong to $N^{\mu\nu}$, and a term $k^\mu k^\nu(q\cdot w)$ with a vector operator $w$ is of the form~\eqref{eq:vertexq} with $A^{\nu\rho}=k^\nu w^\rho-w^\nu k^\rho$ up to a term proportional to $k\cdot q$. A symmetric part of $A^{\nu\rho}$ would leave a pole in the gauge variation, which cannot be canceled.

The invariance under the shift of the polarization requires $q_\mu\xi_\nu\cM^{\mu\nu}=0$ for $\xi\cdot q=0$. From eq.~\eqref{eq:vertexq} we have $q_\mu\xi_\nu\cV_i^{\mu\nu}/(k_i\cdot q)=\xi\cdot k_i+\tfrac12\xi_\nu A_i^{\nu\rho}q_\rho+O(q^2)$. At order $q^0$, we recover eq.~\eqref{eq:gaugecond}. At order $q^1$, we find
\begin{equation}
\sum_i\Big[(\xi\cdot k_i)\,q_\rho\frac{\partial}{\partial k_{i\rho}}+\frac12\,\xi_\nu A_i^{\nu\rho}q_\rho\Big]\cM_N+q_\mu\xi_\nu R^{\mu\nu}=0,
\label{eq:order1}
\end{equation}
where
\begin{equation}
R^{\mu\nu}\equiv N^{\mu\nu}(0)+\sum_i\sum_{n'\neq n_i}\frac{2V^{\mu\nu}_{n_in'}(k_i,0)}{\Delta_{n'n_i}}\,\cM_N^{(i:n')}
\label{eq:Rdef}
\end{equation}
is the sum of all the terms of eq.~\eqref{eq:decomp} that are regular at $q=0$. This is the argument of Low~\cite{Low:1958sn}, in the form given for gravitons in refs.~\cite{Gross:1968in,Bern:2014vva}, with the only difference that the regular part contains the transition terms~\eqref{eq:mixingterm}. Since eq.~\eqref{eq:order1} is linear in $q$ and in $\xi$ and holds for all null $q$ and all $\xi$ with $\xi\cdot q=0$, it determines $R^{\mu\nu}$ up to a multiple of $\eta^{\mu\nu}$,
\begin{equation}
R^{\mu\nu}=-\sum_i\Big[k_i^\nu\frac{\partial}{\partial k_{i\mu}}+\frac12A_i^{\nu\mu}\Big]\cM_N+c\,\eta^{\mu\nu}.
\label{eq:Rsolution}
\end{equation}
This equation has two consequences.

First, the sum of the regular terms is expressed through the amplitude $\cM_N$ of the same levels. The amplitudes $\cM_N^{(i:n')}$ of the other levels, which appear in the individual terms of eq.~\eqref{eq:Rdef}, cancel in the sum. The example of eq.~\eqref{eq:redefvertex} is a special case.

Second, $R^{\mu\nu}$ is symmetric, and the antisymmetric part of the right-hand side of eq.~\eqref{eq:Rsolution} must vanish. With the orbital part of the Lorentz generator of the particle $i$,
\begin{equation}
L_i^{\mu\nu}=-\ii\Big(k_i^\mu\frac{\partial}{\partial k_{i\nu}}-k_i^\nu\frac{\partial}{\partial k_{i\mu}}\Big),
\end{equation}
this condition reads $\sum_i(L_i^{\mu\nu}-\ii A_i^{\mu\nu})\cM_N=0$. The Lorentz invariance of $\cM_N$ is the statement $\sum_i(L_i^{\mu\nu}+\Sigma_i^{\mu\nu})\cM_N=0$, where $\Sigma_i^{\mu\nu}$ is the spin part of the generator acting on the particle $i$. Therefore
\begin{equation}
A_i^{\mu\nu}=\ii\,\Sigma_i^{\mu\nu}+a_i^{\mu\nu},\qquad \sum_ia_i^{\mu\nu}\cM_N=0 .
\label{eq:Asolution}
\end{equation}
We show in section~\ref{sec:vertexq} that $a_i^{\mu\nu}=0$. Inserting eqs.~\eqref{eq:Rsolution} and~\eqref{eq:Asolution} into eq.~\eqref{eq:decomp}, we obtain the subleading soft theorem~\cite{Cachazo:2014fwa,Bern:2014vva},
\begin{equation}
\cM_{N+1}=\big(S^{(0)}+S^{(1)}\big)\cM_N+O(\omega),\qquad
S^{(1)}=-\frac{\ii\kappa}{2}\sum_i\frac{\varepsilon_{\mu\nu}k_i^\mu q_\rho J_i^{\rho\nu}}{k_i\cdot q},
\label{eq:subleading}
\end{equation}
where $S^{(0)}$ is the factor of eq.~\eqref{eq:softG} with $G=\mathbf{1}$ and $J_i^{\mu\nu}=L_i^{\mu\nu}+\Sigma_i^{\mu\nu}$. For external states of arbitrary mass and spin in a generic theory, eq.~\eqref{eq:subleading} was derived in refs.~\cite{Sen:2017xjn,Sen:2017nim,Laddha:2017ygw} from the covariance of the effective action, and in ref.~\cite{Falkowski:2020aso} from on-shell recursion relations. For massless particles in four dimensions, local operators do not modify it~\cite{Elvang:2016qvq}.

The point that we wish to emphasize is the following. The generator $J_i^{\mu\nu}$ acts within the irreducible representation of the particle $i$. It does not change the level. Hence, through order $\omega^0$, the amplitude for the emission of a soft graviton contains no admixture of the amplitudes of other levels, whatever the transition vertices $V^{\mu\nu}_{nn'}$ are. In appendix~\ref{app:toy} we verify eq.~\eqref{eq:subleading} numerically in a model with two levels of different mass that are coupled to each other.

\subsection{The vertex at first order in the soft momentum}
\label{sec:vertexq}

Equation~\eqref{eq:Asolution} states that, on a degenerate block, the part of the vertex that is linear in $q$ is the spin part of the Lorentz generator, up to a matrix $a_i^{\mu\nu}$ whose sum over the external particles annihilates every amplitude. Such a matrix would be an additional conserved charge. It is an antisymmetric tensor that acts on the spin and internal labels only, and in a theory with a non-trivial $S$-matrix no such conserved charge exists besides the generators of the Poincar\'e group~\cite{Coleman:1967ad}. More directly, the condition $\sum_ia_i^{\mu\nu}(k_i)\cM_N=0$ must hold for all momenta, and for generic kinematics the matrices $a_i^{\mu\nu}(k_i)$ of the different particles are independent. Hence $a_i^{\mu\nu}=0$, and
\begin{equation}
\cV_i^{\mu\nu}(k,q)=k^\mu k^\nu+\frac{\ii}{2}\big(k^\mu\Sigma_i^{\nu\rho}+k^\nu\Sigma_i^{\mu\rho}\big)q_\rho+O(q^2).
\label{eq:vertexfixed}
\end{equation}
The vertex through first order in $q$ is fixed by the generators of translations and of Lorentz transformations of the one-particle states. It is diagonal in the levels and contains no free parameter.

It is instructive to list the couplings that are excluded by this result. We denote by $S^\mu$ the spin vector of a massive level, with $S\cdot k=0$.
\begin{itemize}
\item[(a)] An anomalous gravitomagnetic moment, $a^{\mu\nu}=\ii\,b\,\Sigma^{\mu\nu}$, which would change the coefficient of the coupling of the spin to the gravitomagnetic field.
\item[(b)] A gravitoelectric dipole moment, $a^{\nu\rho}=\ii\,(c/M)(k^\nu S^\rho-k^\rho S^\nu)$. Up to a term proportional to $k\cdot q$ it gives the vertex $\ii\,(c/M)\,k^\mu k^\nu(q\cdot S)$. In the rest frame this is a mass dipole proportional to $c\,\bS$, that is, a displacement of the center of energy along the spin, and in a static potential it gives a coupling proportional to $\bS\cdot\nabla\Phi$.
\item[(c)] A transition dipole within a degenerate multiplet, $a^{\nu\rho}=\ii\,(k^\nu d^\rho-k^\rho d^\nu)/M$, where $d^\mu$ is a vector operator that connects different members of the multiplet, for example degenerate states of opposite parity.
\end{itemize}
In all three cases the condition $\sum_ia_i^{\mu\nu}\cM_N=0$ is not a consequence of conservation of angular momentum, and the couplings must vanish. The cases (a) and (b) are the relations of Kobzarev and Okun~\cite{Kobzarev:1962wt}, which are discussed as a manifestation of the equivalence principle, also for higher spin, in refs.~\cite{Teryaev:1999su,Teryaev:2016edw,Silenko:2006er}. They were originally derived from the conservation laws of the energy-momentum tensor. In the on-shell formalism the inconsistency of a dipole coupling was shown in ref.~\cite{Falkowski:2020aso}. The case (c) is their extension to a degenerate multiplet. Experiments search for a coupling of type (b) by comparing the free fall of atoms with different spin orientations~\cite{Tarallo:2014oaa,Duan:2015zmf}. For the graviton it is excluded, and a signal would indicate another interaction.

In the rest frame of a massive level the boost components of the spin generator vanish, $\Sigma^{0j}=0$, and $\Sigma^{jl}=\epsilon^{jlm}S_m$. Equation~\eqref{eq:vertexfixed} then shows that the energy density has no term linear in $q$, so that the mass dipole of every level vanishes when the origin is at its center of energy, and that the term in the momentum density that is linear in $q$ is given by the spin.

\subsection{Scope of the soft expansion and loop corrections}
\label{sec:loops}

We use the soft expansion only through the leading and subleading terms in eq.~\eqref{eq:subleading}. The remainder is not fixed by those two terms. Curvature couplings are allowed interactions that depend on the structure of the system, but no higher-order soft theorem is needed for the matching through dipole order. In particular, a regular local curvature vertex contains two derivatives of the field. This derivative counting should not be inferred from the fixed-gap identity~\eqref{eq:RiemannForm} alone.

The status of eq.~\eqref{eq:subleading} beyond tree level depends on the spacetime dimension. The all-orders proof in ref.~\cite{Sen:2017nim} assumes amplitudes free of infrared divergences, which requires five or more non-compact dimensions. In four dimensions the subleading soft factor is infrared divergent~\cite{Bern:2014oka,He:2014bga}. For infrared-finite amplitudes the expansion contains a term proportional to $\ln\omega$, which dominates over the term of order $\omega^0$. Its coefficient is universal and is expressed through the momenta of the hard particles~\cite{Laddha:2018myi,Sahoo:2018lxl,Krishna:2023fxg}, so that it multiplies the amplitude of the same levels. We do not use eq.~\eqref{eq:subleading} beyond tree level.

\subsection{On-shell information and the gradient expansion}
\label{sec:softsummary}

The results needed for the matching are the following.
\begin{itemize}
\item[(S1)] The leading soft pole fixes the gravitational charge on each level and on every connected degenerate multiplet, eq.~\eqref{eq:Guniversal}. It does not determine a zero-transfer vertex between unequal masses.
\item[(S2)] On a degenerate block the vertex linear in $q$ is fixed by the Lorentz generator, eq.~\eqref{eq:vertexfixed}. For the complete emission amplitude the regular transition and contact contributions combine into the amplitude of the original external levels through order $\omega^0$, eq.~\eqref{eq:subleading}; the individual unequal-mass vertices need not vanish.
\item[(S3)] The remaining information on transitions between unequal masses is supplied by the on-shell transition residue of section~\ref{sec:pole}. It obeys $s_n+s_{n'}\ge2$ and admits the curvature representation~\eqref{eq:RiemannForm}. Its interpretation as a tidal interaction uses the separate matching assumptions of section~\ref{sec:tchannel}.
\end{itemize}

For an external line that retains its mass, the adjacent propagator contributes $1/(k\cdot q)$. A vertex of order $q^0$ or $q^1$ therefore contributes at order $\omega^{-1}$ or $\omega^0$ to the amplitude, respectively. This is the correspondence shown in table~\ref{tab:softgradient}. It is not a prescription for substituting a static off-shell momentum into an on-shell soft factor. In the unequal-mass sector the propagator is regular at $q=0$, and the matching must instead use the exchange residue and its controlled continuation.

\begin{table}[t]
\centering
\begin{tabular}{llll}
\toprule
Soft order & Vertex order & Static matching data & On-shell operator\\
\midrule
leading, $\omega^{-1}$ & $q^0$ & value of $\Phi$ & $P^\mu$\\
subleading, $\omega^0$ & $q^1$ & gradient of $\Phi$, rotation & $J^{\mu\nu}$\\
\bottomrule
\end{tabular}
\caption{Leading and subleading information for an external line with an unchanged mass, including a degenerate multiplet. The static interpretation requires the matching assumptions of section~\ref{sec:assumptions}. Unequal-mass transitions are constrained by separate transition residues; curvature coefficients are not determined by this table.}
\label{tab:softgradient}
\end{table}

A metric perturbation that is constant or linear in the coordinates has zero linearized curvature and can locally be removed by a coordinate transformation. The relation of leading and subleading soft identities to such transformations is discussed in refs.~\cite{Mirbabayi:2016xvc,DeLuca:2024cjl}. Nevertheless, the on-shell statements (S1)--(S3) do not identify an off-shell gravitational mass matrix unless a matching prescription is specified. Section~\ref{sec:invariant} provides this prescription for the test family~\eqref{eq:family} and shows that the constrained quantity is the invariant combination of the potential and momentum-dependent terms.

The following two sections take a different route. A local conserved stress tensor provides both the charges and the off-shell matrix elements, and it is sufficient by itself to derive the laboratory Hamiltonian in a uniform field. We use this additional hypothesis to construct an explicit Hamiltonian and to check its agreement with the invariant matching. We do not use it to complete a proof based on the soft theorems alone.

\section{Constraints from a local conserved source}
\label{sec:source}

We now use the additional source hypothesis of section~\ref{sec:assumptions}. Identifying the gravitational source with a local conserved stress tensor whose charges are the Poincar\'e generators is a stronger assumption than specifying its on-shell soft residues. This section and section~\ref{sec:matching} give an independent derivation under this hypothesis, and the agreement with the amplitude constraints is a consistency check.

\subsection{Matrix elements of the source and charges}

We assume that the graviton couples to a local, symmetric and conserved stress tensor $T^{\mu\nu}$. In this section $p$ and $p'$ denote physical momenta. We normalize the one-particle states as
\begin{equation}
\langle n',\bm{p}',\sigma'|n,\bm{p},\sigma\rangle=2E_{\bm{p}}\,(2\pi)^3\delta^3(\bm{p}-\bm{p}')\,\delta_{n'n}\delta_{\sigma'\sigma},
\end{equation}
and consider the matrix elements of the source,
\begin{equation}
T^{\mu\nu}_{n'n}(p',p)=\langle n',\bm{p}'|T^{\mu\nu}(0)|n,\bm{p}\rangle,\qquad q=p'-p,\qquad P=p+p',
\label{eq:vertex}
\end{equation}
where the spin labels are suppressed. These matrix elements are defined for all values of $q^2$ and provide an off-shell continuation of the vertices of eq.~\eqref{eq:decomp}, to which they reduce on the mass shell of the graviton up to the normalization factor $2$. Translation covariance and the conservation law $\partial_\mu T^{\mu\nu}=0$ give the Ward identity
\begin{equation}
q_\mu T^{\mu\nu}_{n'n}(p',p)=0 .
\label{eq:Ward}
\end{equation}
The charges of the source are the generators of translations,
\begin{equation}
P^\mu=\int\dd^3x\,T^{0\mu}(0,\bx).
\label{eq:charges}
\end{equation}
Furthermore, the conservation law implies $\int\dd^3x\,T^{ij}=\tfrac12\,\dd^2/\dd t^2\int\dd^3x\,x^ix^jT^{00}$, whose matrix elements between states at rest with equal energy vanish. Taking the matrix elements of these relations between states of the same mass at rest, and using Lorentz covariance, we obtain
\begin{equation}
T^{\mu\nu}_{n'n}(p,p)=2p^\mu p^\nu\,\delta_{n'n}\delta_{\sigma'\sigma}\qquad(M_{n'}=M_n).
\label{eq:forward}
\end{equation}
The charge relation therefore gives $G=\mathbf{1}$ for the gravitational charge matrix of eq.~\eqref{eq:softG}, in agreement with eq.~\eqref{eq:Guniversal}. In the same way, the Lorentz generators $J^{\mu\nu}=\int\dd^3x\,(x^\mu T^{0\nu}-x^\nu T^{0\mu})$ fix the term linear in $q$ to be the total angular momentum, which is equivalent to $A(0)=1$ and $B(0)=0$ for the gravitational form factors and to the absence of a gravitoelectric dipole moment~\cite{Kobzarev:1962wt,Teryaev:2016edw}. These constraints have been derived non-perturbatively for arbitrary spin in refs.~\cite{Cotogno:2019xcl,Lorce:2019sbq}. Since a stable level of a composite system is a massive one-particle state, they apply to each level without modification, and they agree with eq.~\eqref{eq:vertexfixed}. Equations~\eqref{eq:Guniversal} and~\eqref{eq:vertexfixed} establish the corresponding on-shell constraints without the local-source hypothesis. Here the charge identities reproduce them independently and supply matrix elements at $q^2\neq0$, to which we now turn.

\subsection{Spinless levels}
\label{sec:spinless}

Parametrizations of the matrix elements of the energy-momentum tensor between states of different mass have been given for spins $\tfrac12$ and $\tfrac32$ in refs.~\cite{Kim:2022bwn,Ozdem:2022zig}. For our purpose the simplest case is sufficient. Let $n$ and $n'$ be two spinless levels with $M_{n'}\neq M_n$, and let
\begin{equation}
\Delta\equiv M_{n'}^2-M_n^2=-P\cdot q .
\end{equation}
The most general symmetric tensor is
\begin{equation}
T^{\mu\nu}_{n'n}=\frac12A\,P^\mu P^\nu+\frac12B\,\big(q^\mu q^\nu-q^2\eta^{\mu\nu}\big)+\frac12C\,\big(P^\mu q^\nu+P^\nu q^\mu\big)+D\,\eta^{\mu\nu},
\label{eq:general}
\end{equation}
where $A$, $B$, $C$ and $D$ are functions of $q^2$. The Ward identity~\eqref{eq:Ward} gives
\begin{equation}
q_\mu T^{\mu\nu}_{n'n}=\frac12\big(-A\Delta+Cq^2\big)P^\nu+\Big(-\frac12C\Delta+D\Big)q^\nu=0,
\end{equation}
and since $P$ and $q$ are linearly independent for $\Delta\neq0$, we find
\begin{equation}
A=\frac{C\,q^2}{\Delta},\qquad D=\frac{C\,\Delta}{2}.
\label{eq:solution}
\end{equation}
For $\Delta=0$ the same identity gives $C=D=0$ instead, and eq.~\eqref{eq:general} reduces to the familiar form with the form factors $A$ and $B$ and with $A(0)=1$ from eq.~\eqref{eq:forward}~\cite{Pagels:1966zza}.

We now contract with the polarization $\varepsilon_{\mu\nu}$ of a plane wave $h_{\mu\nu}=\varepsilon_{\mu\nu}e^{\ii q\cdot x}$, which need not be on shell. The linearized Ricci tensor and Ricci scalar of this wave are
\begin{equation}
R_{\mu\nu}=\frac12\big(q^2\varepsilon_{\mu\nu}+q_\mu q_\nu\,\varepsilon^\rho{}_\rho-q_\mu q^\rho\varepsilon_{\rho\nu}-q_\nu q^\rho\varepsilon_{\rho\mu}\big),\qquad
R=q^2\varepsilon^\rho{}_\rho-q^\mu q^\nu\varepsilon_{\mu\nu},
\label{eq:Ricci}
\end{equation}
where the factor $e^{\ii q\cdot x}$ is omitted. Inserting eq.~\eqref{eq:solution} into eq.~\eqref{eq:general}, we find
\begin{equation}
\varepsilon_{\mu\nu}T^{\mu\nu}_{n'n}=\frac{C}{\Delta}\,R_{\mu\nu}P^\mu P^\nu-\frac{B}{2}\,R .
\label{eq:RicciForm}
\end{equation}
Thus, the complete one-graviton vertex between two spinless levels of different mass is a coupling to the linearized Ricci tensor. For comparison, ref.~\cite{Draper:2020knh} parametrized the non-minimal vertex of a single scalar species by Ricci form factors. The independent sign symmetry of each scalar field in that model excludes transitions between the two species. For the unequal-mass transition considered here, the conservation law brings the entire vertex into the Ricci form. This result has three consequences.

First, there is no transition monopole. The coefficient of $P^\mu P^\nu$ is $A=Cq^2/\Delta$, which vanishes at $q^2=0$ if $C$ is regular there. This is the gravitational analogue of the vanishing of the transition charge for a conserved current, $\langle n'|J^\mu|n\rangle=F_1P^\mu+F_2q^\mu$ with $F_1=F_2q^2/\Delta$.

Second, the vertex vanishes for an on-shell graviton, for which $q^2=0$, $q^\mu\varepsilon_{\mu\nu}=0$ and $\varepsilon^\rho{}_\rho=0$, so that $R_{\mu\nu}=0$. This reproduces the result $a=0$ of section~\ref{sec:selection}.

Third, in a static external field the first-order transition amplitude is evaluated on the energy shell, $q^0=0$, and $R_{\mu\nu}$ in eq.~\eqref{eq:RicciForm} is the Ricci tensor of the background. By the linearized Einstein equation it is proportional to the stress tensor of the source. For form factors regular at $q^2=0$, the massless exchange pole therefore cancels, in agreement with section~\ref{sec:tchannel}. In the local derivative expansion the resulting terms are supported on the source and vanish outside it. The same holds for the term proportional to $B$ in the diagonal vertex, which is the $D$-term of the literature on gravitational form factors.

We stress that eq.~\eqref{eq:RicciForm} refers to the matrix elements between on-shell states of the system. The matrix elements of the interaction Hamiltonian between states of different energy, which enter at higher orders of time-dependent perturbation theory, are evaluated with $q^0=E_{n'}-E_n\neq0$ for a static field and are not zero in general. They depend on the gauge, as we have seen in section~\ref{sec:softnondeg}. We return to this point in section~\ref{sec:rep}.

\subsection{Levels with spin}
\label{sec:spin}

For transitions between levels with spin we have no classification comparable to eq.~\eqref{eq:RicciForm}. The identity~\eqref{eq:lemma} holds for any $q\neq0$, also off the mass shell of the graviton. A conserved vertex therefore depends on the polarization only through the Riemann tensor, $\varepsilon_{\mu\nu}T^{\mu\nu}=-2T^{\mu\sigma}\ell^\nu\ell^\rho R_{\mu\nu\rho\sigma}[\varepsilon]$, and for $\Delta\neq0$ one can take $\ell=-P/\Delta$. This statement is a consequence of the conservation law alone. It does not imply that the coefficient of the Riemann tensor is local or regular as $q\to0$, and we do not use it to classify the transition couplings. On the mass shell of the graviton it reduces to eq.~\eqref{eq:RiemannForm}, and the selection rule~\eqref{eq:selection} applies.

The two routes should be kept separate. The on-shell analysis fixes the universal monopole and dipole data of each level and degenerate multiplet and constrains the pole residues of mass-changing amplitudes. Under the additional source hypothesis, the charge identities reproduce these universal data, and the Ward identity supplies further off-shell information. A curvature representation alone does not classify every regular low-momentum operator for arbitrary spin. The implications for the specified test Hamiltonian are discussed in section~\ref{sec:invariant}.

\section{Derivation of the laboratory Hamiltonian}
\label{sec:matching}

In this section, we derive the laboratory Hamiltonian directly from the local-source coupling and the Poincar\'e algebra. This derivation does not use any result of the soft expansion. It provides an explicit Hamiltonian and an independent realization of the invariant constraints discussed in section~\ref{sec:invariant}.

\subsection{Background coupling and the proper reference frame}

From this section on we restore $\hbar$ and $c$. We take $x^0=ct$, so that $T^{00}$ is the energy density, $T^{0i}/c$ the momentum density and $T^{ij}$ the stress. The coupling to $T^{\mu\nu}$ gives the Hamiltonian of the matter system to first order in a weak classical background $g_{\mu\nu}=\eta_{\mu\nu}+h_{\mu\nu}$,
\begin{equation}
H[h]=H-\frac12\int\dd^3x\,h_{\mu\nu}(\bx)\,T^{\mu\nu}(\bx).
\label{eq:Hh}
\end{equation}
We now evaluate eq.~\eqref{eq:Hh} in the frame in which a laboratory experiment is described.

Consider an observer with proper acceleration $\ba$ whose spatial axes rotate with angular velocity $\bom$ relative to Fermi-Walker transport. In the coordinates of the proper reference frame of this observer the metric is~\cite{Misner:1973prb,Ni:1978zz}
\begin{align}
g_{00}&=-\Big(1+\frac{\ba\cdot\bx}{c^2}\Big)^2+\frac{(\bom\times\bx)^2}{c^2}-R_{0l0m}x^lx^m,\nonumber\\
g_{0i}&=\frac{(\bom\times\bx)_i}{c}-\frac23R_{0lim}x^lx^m,\qquad
g_{ij}=\delta_{ij}-\frac13R_{iljm}x^lx^m,
\label{eq:FWmetric}
\end{align}
up to terms of third order in $\bx$. We call these coordinates the Fermi-Walker frame. They should be distinguished from the Fermi normal coordinates along the worldline of a freely falling atom, for which $\ba=0$. A laboratory at rest in a static gravitational potential $\Phi$ has $\ba=\nabla\Phi$. Inserting the first-order part of eq.~\eqref{eq:FWmetric} into eq.~\eqref{eq:Hh}, we obtain
\begin{equation}
\Hlab=H+\ba\cdot\bK-\bom\cdot\bJ+\Htid,
\label{eq:Hlab}
\end{equation}
with
\begin{equation}
\bK=\frac{1}{c^2}\int\dd^3x\,\bx\,T^{00},\qquad
J_i=\frac1c\,\epsilon_{ijk}\int\dd^3x\,x_jT^{0k}.
\label{eq:KJ}
\end{equation}
These are the generators of boosts, evaluated at $t=0$, and of rotations of the isolated system. The spatial components $h_{ij}$ vanish at this order, so that the stress $T^{ij}$ does not appear. The tidal term $\Htid$ collects the contributions of the curvature and is discussed in section~\ref{sec:tidal}. In flat spacetime eq.~\eqref{eq:Hlab} with $\Htid=0$ is the generator of the evolution in the proper time of the observer and holds for any matter system. The statement that the same expression describes a laboratory at rest in a gravitational field is the equivalence principle for the matter theory, which follows here from eq.~\eqref{eq:Hh}.

Equation~\eqref{eq:Hlab} contains the same couplings to the momentum and the angular momentum as (S1) and (S2), but its derivation uses eq.~\eqref{eq:Hh} and the identification of the moments of the source with the Poincar\'e charges. The soft theorems are not used in this argument, and they do not by themselves establish this off-shell identification. The agreement is a consistency check. For a hydrogen-like atom in Fermi-Walker coordinates, a Hamiltonian of this structure and the corresponding center-of-mass coordinates were derived in ref.~\cite{Janson:2024hdt}.

\subsection{The mass operator in the laboratory frame}
\label{sec:massop}

The Poincar\'e generators satisfy
\begin{align}
[K_i,H]&=\ii\hbar P_i, & [K_i,P_j]&=\ii\hbar\,\delta_{ij}\frac{H}{c^2},\nonumber\\
[J_i,K_j]&=\ii\hbar\,\epsilon_{ijk}K_k, & [K_i,K_j]&=-\frac{\ii\hbar}{c^2}\,\epsilon_{ijk}J_k .
\label{eq:algebra}
\end{align}
\begin{proposition}
\label{prop:mass}
For arbitrary time-dependent $\ba(t)$ and $\bom(t)$,
\begin{equation}
\big[H+\ba(t)\cdot\bK-\bom(t)\cdot\bJ,\;\hat{M}\big]=0,
\label{eq:commM}
\end{equation}
where $\hat{M}$ is the mass operator of eq.~\eqref{eq:Mop}.
\end{proposition}
\noindent The proof is immediate. From eq.~\eqref{eq:algebra} we have $[K_i,H^2]=2\ii\hbar HP_i$ and $[K_i,\bP^2]=2\ii\hbar P_iH/c^2$, so that $[K_i,\hat{M}^2]=0$. The rotations commute with $H$ and $\bP^2$. Since $\hat{M}$ is the positive square root of $\hat{M}^2$, eq.~\eqref{eq:commM} follows.

Proposition~\ref{prop:mass} states that the acceleration and the rotation of the frame do not induce transitions between levels of different mass. This holds for arbitrary motion of the center of mass and requires no adiabaticity. Of course, the real interactions with the support, with trapping fields or with lasers are additional terms in the Hamiltonian. For example, an electric field that holds an ion against gravity mixes its levels through the Stark effect. Such effects are not gravitational couplings of the internal degrees of freedom.

To compare with the test theory we need the explicit form of $\bK$. In an irreducible representation with mass $M_n$ and spin $s_n$, the generators can be written as~\cite{Newton:1949cq,Foldy:1956kvk}
\begin{equation}
H=\sqrt{M_n^2c^4+c^2\bP^2},\qquad \bJ=\bX\times\bP+\bS,\qquad
\bK=\frac{1}{2c^2}\{\bX,H\}-\frac{\bS\times\bP}{H+M_nc^2},
\label{eq:NW}
\end{equation}
where $[X_i,P_j]=\ii\hbar\delta_{ij}$ and $\bS$ is the spin operator. We have verified that eq.~\eqref{eq:NW} satisfies eq.~\eqref{eq:algebra} (appendix~\ref{app:poincare}). Since the generators act within each level, replacing $M_n$ in eq.~\eqref{eq:NW} by $\hat{M}=\sum_nM_n\Pi_n$ gives the generators on the full space of stable levels. The operator $\hat{M}$ commutes with $\bX$, $\bP$ and $\bS$. Expanding in $1/c^2$ we find
\begin{align}
\Hlab={}&\hat{M}c^2+\frac{\bP^2}{2\hat{M}}+\hat{M}\,\ba\cdot\bX-\bom\cdot(\bX\times\bP+\bS)\nonumber\\
&+\frac{1}{4\hat{M}c^2}\{\ba\cdot\bX,\bP^2\}+\frac{1}{2\hat{M}c^2}\bS\cdot(\ba\times\bP)-\frac{\bP^4}{8\hat{M}^3c^2}+\Htid+O(c^{-4}).
\label{eq:Hexp}
\end{align}
For a spin-$\tfrac12$ particle this agrees with the Hamiltonian of a Dirac particle in an accelerated and rotating frame~\cite{Hehl:1990nf}.

The time coordinate of the Fermi-Walker frame is the proper time at its origin, which corresponds to $\Phi_0=0$. Relative to a reference time for which the potential at the origin is $\Phi_0$, the Hamiltonian acquires the additional first-order term $\Phi_0H/c^2$, which contributes $\hat M\Phi_0$ at the order of eq.~\eqref{eq:Htest}. With this normalization of time understood, we can compare eq.~\eqref{eq:Hexp} with the test Hamiltonian for $\Phi(\bX)=\Phi_0+\ba\cdot\bX$. The same operator $\hat{M}=m\mathbf{1}+\Hint/c^2$ appears in the rest energy, in the kinetic term and in the potential term. Hence
\begin{equation}
\Mr=\Mi=\Mg=\hat{M}=m\mathbf{1}+\frac{\Hint}{c^2},
\label{eq:QEPresult}
\end{equation}
that is, $r_1=r_2=1$ and $r=0$ in eq.~\eqref{eq:rmatrix}. This equality holds in the stated representation under the local-source hypothesis, at first order in the field and below tidal order; it is not inferred from the soft factors alone. The term $\hat{M}\ba\cdot\bX$ contains the redshift coupling $\Hint\,\ba\cdot\bX/c^2$ of the internal energy~\cite{Zych:2011hu,Pikovski:2013qwa}, with no off-diagonal element. Three remarks are in order.

First, eq.~\eqref{eq:QEPresult} is a statement in a definite representation. The internal states are the eigenstates of the mass operator, and $\bX$ is the Newton-Wigner position operator of the composite system. For spinless levels it coincides with the center of energy $c^2\{\bK,H^{-1}\}/2$. It differs from the center of mass defined with the rest masses of the constituents by terms of relative order $E_{\rm int}/(mc^2)$, which have been worked out for atoms in refs.~\cite{Krajcik:1974nv,Sonnleitner:2018xbk,Schwartz:2019qqg,Janson:2024hdt}. In another representation the Hamiltonian has a different form, as we discuss in section~\ref{sec:rep}.

Second, the terms that contain the spin are universal. The coupling $-\bom\cdot\bS$~\cite{Mashhoon:1988zz} and the spin-orbit coupling $\bS\cdot(\ba\times\bP)/(2\hat{M}c^2)$~\cite{Hehl:1990nf} have no anomalous coefficients, and $\bS$ acts within the spin multiplet of each level. There is no term proportional to $\bS\cdot\ba$. A boost generator that is displaced by a multiple of the spin, $\bK+c_{\rm d}\bS$, does not satisfy the last relation of eq.~\eqref{eq:algebra} unless $c_{\rm d}=0$ (appendix~\ref{app:poincare}). This is the laboratory-frame form of the result of section~\ref{sec:vertexq}. The spin of a level is the total internal angular momentum of the composite system.

Third, there is no gravitational analogue of the Stark effect for a freely moving system. A uniform electric field couples to the dipole operator, which is not a symmetry generator and connects different levels. A uniform gravitational field couples to the boost generator, which commutes with $\hat{M}$.

\subsection{Tidal order}
\label{sec:tidal}

The curvature terms of eq.~\eqref{eq:FWmetric} give
\begin{equation}
\Htid=\frac{c^2}{2}R_{0l0m}\,I^{lm}+\cdots,\qquad I^{lm}=\frac{1}{c^2}\int\dd^3x\,x^lx^mT^{00},
\label{eq:Htid}
\end{equation}
where the omitted terms involve the moments of $T^{0i}$ and $T^{ij}$ and are suppressed by powers of the internal velocities for a non-relativistic system~\cite{Parker:1980kw}. In a static field $c^2R_{0l0m}=\partial_l\partial_m\Phi$. Outside the source $\nabla^2\Phi=0$ and only the traceless part $Q^{lm}=I^{lm}-\tfrac13\delta^{lm}I^{kk}$ contributes,
\begin{equation}
\Htid=\frac12\,Q^{lm}\,\partial_l\partial_m\Phi+\cdots .
\label{eq:quadrupole}
\end{equation}
The mass quadrupole $Q^{lm}$ is an operator on $\cH_{\rm int}$ that does not commute with $\Hint$ in general. This is the first term in the multipole expansion of the local source that depends on the internal structure and can connect levels of different mass, consistent with the curvature representation in (S3). Such tidal transitions are allowed finite-size effects in ordinary gravity and are not, by themselves, violations of the equivalence principle. The matrix elements of $Q^{lm}$ between two levels are non-zero only if $s_n+s_{n'}\ge2$, which is the selection rule~\eqref{eq:selection}. The trace part of $I^{lm}$ couples to $\nabla^2\Phi$, that is, to the Ricci tensor, and vanishes outside the source, in agreement with eq.~\eqref{eq:RicciForm}. The tidal polarizabilities characterize the response at second order in the curvature and are obtained after integrating out the excited levels~\cite{Goldberger:2004jt}. The situation is summarized in table~\ref{tab:multipole}.

\begin{table}[t]
\centering
\begin{tabular}{llll}
\toprule
Field quantity & Operator & Status & Mixing of levels\\
\midrule
$\Phi_0$ & $\hat{M}$ & universal & no\\
$\nabla\Phi$ & $\bK$ & universal & no\\
$\bom$ & $\bJ$ & universal & no\\
$\partial_l\partial_m\Phi$ (traceless) & $Q^{lm}$ & structure dependent & yes (selection rules)\\
$\nabla^2\Phi$ & ${\rm tr}\,I$, $D$-term & zero outside the source & contact term\\
$(\partial\partial\Phi)^2$ & polarizabilities & second order & --\\
\bottomrule
\end{tabular}
\caption{Coupling of a composite system to a static gravitational field, organized by the number of derivatives of the potential. The first three rows are fixed by the Poincar\'e generators of the isolated system and correspond to the first two rows of table~\ref{tab:softgradient}.}
\label{tab:multipole}
\end{table}

The tidal coupling sets the size of the largest off-diagonal effect that gravity itself can produce. To leading order in the internal energy divided by $mc^2$, eq.~\eqref{eq:quadrupole} gives the tidal acceleration operator $-(2m)^{-1}Q^{lm}\partial_i\partial_l\partial_m\Phi$. In the field of the Earth its magnitude relative to $g$ is $3|Q^{zz}|/(mR_\oplus^2)$. For an atomic mass quadrupole of order $m_ea_0^2$ and the mass of $^{87}$Rb this is about $10^{-39}$. For the hyperfine levels of an $S$ state the electronic quadrupole vanishes and the nuclear quadrupole gives about $10^{-44}$. Moreover, for a tidal field that is axially symmetric about the quantization axis, the matrix element between $|F=1,m_F=0\rangle$ and $|F=2,m_F=0\rangle$ vanishes by the selection rule of the rank-two tensor. Hence, the off-diagonal parameter induced by gravity in the experiment of ref.~\cite{Rosi:2017ieh} is smaller than the reported bound by more than thirty-five orders of magnitude.

\section{Representation independence and on-shell matching}
\label{sec:rep}

The result~\eqref{eq:QEPresult} was obtained in the Fermi-Walker frame. Calculations for composite systems are often performed in the Newtonian gauge, where the Hamiltonian has a different form. In sections~\ref{sec:unitary} and~\ref{sec:invariant} we relate the two descriptions and identify the invariant data. The matching to the amplitude analysis, which is the central step, is given in section~\ref{sec:invariant}. It constrains a combination of the potential and momentum-dependent terms and does not identify an off-shell vertex with $\Mg$.

\subsection{Newtonian gauge and the virial term}

In the Newtonian gauge the metric of a weak static field is
\begin{equation}
\dd s^2=-\Big(1+\frac{2\Phi}{c^2}\Big)c^2\dd t^2+\Big(1-\frac{2\gamma\Phi}{c^2}\Big)\dd\bx^2,
\label{eq:Newtonian}
\end{equation}
where $\gamma$ is the post-Newtonian parameter, with $\gamma=1$ in general relativity. Equation~\eqref{eq:Hh} gives
\begin{equation}
\HN=H+\frac{1}{c^2}\int\dd^3x\,\Phi(\bx)\big[T^{00}(\bx)+\gamma\,T^{ii}(\bx)\big].
\label{eq:HN}
\end{equation}
This Hamiltonian contains a stress term that is absent from eq.~\eqref{eq:Hlab}. If a system is at rest at $\bX$ and its size is small compared with the scale over which $\Phi$ varies, the leading term is $\Phi(\bX)\,(\hat{M}+\gamma\Sigma/c^2)$ with
\begin{equation}
\Sigma=\int\dd^3x\,T^{ii}=\frac12\frac{\dd^2}{\dd t^2}I^{kk},
\label{eq:virial}
\end{equation}
where $I^{lm}$ is defined in eq.~\eqref{eq:Htid} and the second equality follows from the conservation law. For a non-relativistic system bound by Coulomb forces, $\Sigma=2K_{\rm int}+U_{\rm int}$, where $K_{\rm int}$ and $U_{\rm int}$ are the internal kinetic and potential energies. The coefficient of the potential for $\gamma=1$ is then $m+(3K_{\rm int}+2U_{\rm int})/c^2$, which is the well-known result of refs.~\cite{Nordtvedt:1994gp,Carlip:1997gf}. The expectation value of $\Sigma$ in a stationary state vanishes by the virial theorem. However, its matrix elements between levels of different energy do not,
\begin{equation}
\langle n'|\Sigma|n\rangle=-\frac{(E_{n'}-E_n)^2}{2\hbar^2}\,\langle n'|I^{kk}|n\rangle .
\label{eq:virialoffdiag}
\end{equation}
If the coefficient of $\Phi(\bX)$ in eq.~\eqref{eq:HN} is identified with the gravitational mass operator, it does not commute with $\Hint$. This is the origin of the claim of refs.~\cite{Lebed:2011np,Lebed:2014oua}. It was clarified in ref.~\cite{Zych:2018bmk} that the gravitational mass of a bound system is given by its total internal energy when the internal dynamics is described in terms of local quantities (see also refs.~\cite{Schwartz:2019qqg,Giulini:2022wyp}). We now give a derivation of this fact that does not depend on the model of the bound system.

\subsection{Unitary equivalence with the Fermi-Walker frame}
\label{sec:unitary}

We define the dilatation and special conformal moments of the momentum density,
\begin{equation}
D=\frac1c\int\dd^3x\,x^iT^{0i},\qquad
C^k=\frac1c\int\dd^3x\,\Big(x^kx^iT^{0i}-\frac12\bx^2\,T^{0k}\Big).
\label{eq:DC}
\end{equation}
The conservation law and the symmetry of $T^{\mu\nu}$ give, after integration by parts,
\begin{equation}
\frac{\ii}{\hbar}[H,D]=\int\dd^3x\,T^{ii},\qquad
\frac{\ii}{\hbar}[H,C^k]=\int\dd^3x\,x^k\,T^{ii}.
\label{eq:DCidentities}
\end{equation}
The derivation is given in appendix~\ref{app:conformal}. The operators $D$ and $C^k$ are in general not conserved and do not generate symmetries. Only their definitions in terms of $T^{0i}$ are used.
\begin{proposition}
\label{prop:unitary}
Let the matter theory have a conserved symmetric stress tensor, and let $\Phi(\bx)=\Phi_0+\ba\cdot\bx$. Then
\begin{equation}
\HN=U_\gamma\,\HFW\,U_\gamma^\dagger+O(\Phi^2),\qquad
\HFW=\Big(1+\frac{\Phi_0}{c^2}\Big)H+\ba\cdot\bK,
\label{eq:equivalence}
\end{equation}
with
\begin{equation}
U_\gamma=\exp\Big[-\frac{\ii\gamma}{\hbar c^2}\big(\Phi_0D+\ba\cdot\bC\big)\Big].
\label{eq:Ugamma}
\end{equation}
\end{proposition}
\noindent To prove this, we insert the linear potential into eq.~\eqref{eq:HN}. The term with $T^{00}$ gives $\Phi_0H/c^2+\ba\cdot\bK$ by eq.~\eqref{eq:KJ}. The term with $T^{ii}$ gives $(\ii\gamma/\hbar c^2)[H,\Phi_0D+\ba\cdot\bC]$ by eq.~\eqref{eq:DCidentities}. Hence $\HN=\HFW+(\ii\gamma/\hbar c^2)[\HFW,\Phi_0D+\ba\cdot\bC]+O(\Phi^2)$, which is the expansion of eq.~\eqref{eq:equivalence} to first order.

The Hamiltonian $\HFW$ is that of the Fermi-Walker frame, eq.~\eqref{eq:Hlab}, up to the constant rescaling of time by $1+\Phi_0/c^2$. Proposition~\ref{prop:unitary} has several consequences.

First, the Hamiltonians in the two coordinate systems are unitarily equivalent at this order, for any matter theory. The transformation $U_\gamma$ acts on the positions of the constituents as
\begin{equation}
\bx\;\to\;\bx-\frac{\gamma}{c^2}\Big[\Phi_0\,\bx+(\ba\cdot\bx)\,\bx-\frac12\bx^2\ba\Big],
\end{equation}
which is the composition of a dilatation and a special conformal transformation. It maps the conformally flat spatial metric of eq.~\eqref{eq:Newtonian} to the flat one of eq.~\eqref{eq:FWmetric}. The virial term is therefore a consequence of the choice of coordinates. In the Newtonian gauge the mass operator is represented by $\hat{M}_{\rm N}=U_\gamma\hat{M}U_\gamma^\dagger$, which commutes with $\HN$ at this order, and the levels are $U_\gamma|n\rangle$.

Second, the parameter $\gamma$ drops out. This is expected, since $\gamma$ measures the spatial curvature, which affects a local experiment only at tidal order.

Third, for a non-relativistic system localized near $\bX$ the generator decomposes as
\begin{equation}
\Phi_0D+\ba\cdot\bC=\Phi(\bX)\,D_{\rm int}+(\ba\times\bX)\cdot\bS_{\rm int}+\ba\cdot\bC_{\rm int}+(\text{center-of-mass terms}),
\label{eq:decomposition}
\end{equation}
where the internal moments are defined with respect to $\bX$ and $\bS_{\rm int}$ is the internal angular momentum. The leading internal part of $U_\gamma$ is $\exp[-\ii\gamma\Phi(\bX)D_{\rm int}/\hbar c^2]$, a rescaling of the internal coordinates that depends on the position of the system. Since $\Phi(\bX)$ does not commute with $\bP$, this transformation also generates the terms in $\HN$ that couple the momentum of the center of mass to internal operators.

\subsection{Predictions do not depend on the representation}

Under a unitary transformation $U(t)$, which may depend on time through the external field, the Hamiltonian and the evolution operator become
\begin{equation}
H'=UHU^\dagger+\ii\hbar\,\dot UU^\dagger,\qquad V'(t_f,t_i)=U(t_f)\,V(t_f,t_i)\,U^\dagger(t_i).
\end{equation}
If the initial state $\rho_i$ and the measured observable $O_f$ are transformed as $\rho_i'=U(t_i)\rho_iU^\dagger(t_i)$ and $O_f'=U(t_f)O_fU^\dagger(t_f)$, all probabilities are unchanged,
\begin{equation}
{\rm Tr}\big(O_f'\,V'\rho_i'\,V'^\dagger\big)={\rm Tr}\big(O_f\,V\rho_i\,V^\dagger\big).
\label{eq:invariance}
\end{equation}
This is exact for any duration of the experiment. As an illustration, consider a pure redshift $H_{\rm red}=(1+f(t))H_0$ and the transformation $U=\exp[-\ii f(t)F]$ with a Hermitian operator $F$. At first order in $f$ the transformed Hamiltonian has the off-diagonal elements $H'_{ab}=\hbar F_{ab}(\dot f+\ii\Omega_{ab}f)$ with $\hbar\Omega_{ab}=E_a-E_b$, and the apparent transition amplitude is a boundary term,
\begin{equation}
-\frac{\ii}{\hbar}\int_{t_i}^{t_f}\dd t\,e^{\ii\Omega_{ab}t}H'_{ab}=-\ii F_{ab}\Big[f(t)\,e^{\ii\Omega_{ab}t}\Big]_{t_i}^{t_f},
\end{equation}
which is canceled by the transformation of the initial and final states.

Equation~\eqref{eq:invariance} is an identity and by itself does not show that a given off-diagonal term is unphysical. The non-trivial statement is proposition~\ref{prop:unitary}, which shows that the Hamiltonian derived in the Newtonian gauge and the one derived in the Fermi-Walker frame are related by a unitary transformation. Together they imply that the off-diagonal elements~\eqref{eq:virialoffdiag} have no observable consequence, provided that the preparation and the detection are described in the same representation. The situation is analogous to the coupling of an atom to the electromagnetic field in the length and velocity gauges, where the matrix elements between the unperturbed levels depend on the gauge and the predictions agree only if the states are transformed consistently~\cite{PhysRevA.36.2763,DiStefano:2018fsc,Stokes:2019upq}.

\subsection{Invariant transition coefficient and matching to the test theory}
\label{sec:invariant}

We now match the long-range amplitude constraints to the test theory. We work at first order in a static weak potential, through dipole order in its gradients, and at the leading retained orders in the internal energy and in the velocity. The universal orbital and spin couplings are assumed to have been separated. We restrict the remaining internal couplings to the following family:
\begin{equation}
H=\hat{M}c^2+\frac{\bP^2}{2\hat{M}}+\Phi(\bX)\,\mathcal{G}+\frac{1}{2M_{\rm ref}c^2}\big\{P_i,\partial_i\Phi(\bX)\big\}\,\cA,
\label{eq:family}
\end{equation}
where $\{\,,\}$ is the anticommutator and $\mathcal{G}$ and $\cA$ are Hermitian operators on $\cH_{\rm int}$. This family is closed under the field-dependent changes of basis introduced below at the stated order. It is not a complete basis of all momentum-dependent or spin-dependent interactions. Curvature operators and terms supported on the source must be treated separately, as specified in section~\ref{sec:assumptions}. In the velocity-dependent term we use the common reference mass $M_{\rm ref}=m$. Replacing $1/\hat{M}$ by $1/M_{\rm ref}$ there changes the Hamiltonian only at higher order in $1/c^2$. The test theory~\eqref{eq:Htest} corresponds to $\mathcal{G}=\Mg$ and $\cA=0$. A change of the internal basis that depends on the field, $U=\exp[-\ii\Phi(\bX)F/\hbar c^2]$ with $F$ Hermitian on $\cH_{\rm int}$, maps eq.~\eqref{eq:family} into a Hamiltonian of the same form with
\begin{equation}
\mathcal{G}\;\to\;\mathcal{G}+\frac{\ii}{\hbar c^2}[\Hint,F],\qquad \cA\;\to\;\cA+F,
\label{eq:transformation}
\end{equation}
up to terms of higher order in $1/c^2$. The second relation follows from $U\bP U^\dagger=\bP+F\nabla\Phi/c^2$. We conclude the following. The diagonal elements of $\mathcal{G}$ in the eigenbasis of $\Hint$ and its elements within degenerate blocks are invariant. The elements of $\mathcal{G}$ between non-degenerate levels can be shifted by an arbitrary amount, since any operator that is off-diagonal between non-degenerate levels is of the form $\ii[\Hint,F]$. The invariant combination is
\begin{equation}
\rinv_{mn}\equiv\frac{1}{M_{\rm ref}}\Big[\mathcal{G}_{mn}-\frac{\ii}{\hbar c^2}(E_m-E_n)\,\cA_{mn}\Big]\qquad(E_m\neq E_n).
\label{eq:rinv}
\end{equation}
The off-diagonal parameter $r$ of eq.~\eqref{eq:rmatrix} coincides with $\rinv$ only in a representation with $\cA=0$. Here, invariance refers to the transformation~\eqref{eq:transformation} within the retained truncation. Under a constant change of the internal basis, the corresponding operator transforms covariantly, and an individual matrix element also depends on the phases chosen for the energy eigenstates. Whether it vanishes does not depend on these phases.

\paragraph{The coefficient measured by a transition amplitude.}
In the first Born approximation, the static transition kernel from $|n,\bm p\rangle$ to $|m,\bm p'\rangle$, after a common plane-wave normalization factor is removed, is
\begin{equation}
\mathcal T^{(1)}_{mn}
=\tilde\Phi(\bm q)\Big[\mathcal{G}_{mn}
+\frac{\ii(\bm p'+\bm p)\cdot\bm q}{2M_{\rm ref}\hbar c^2}\,\cA_{mn}\Big],
\qquad \bm q=\bm p'-\bm p .
\label{eq:Bornfamily}
\end{equation}
Here the Fourier transform is defined with the factor $e^{\ii\bm q\cdot\bX/\hbar}$, so that $\partial_i\Phi$ gives $\ii q_i\tilde\Phi/\hbar$. At the required order, energy conservation in a static field gives
\begin{equation*}
E_n+\frac{\bm p^2}{2M_{\rm ref}}
=E_m+\frac{\bm p'^2}{2M_{\rm ref}},
\qquad
\frac{(\bm p'+\bm p)\cdot\bm q}{2M_{\rm ref}}=E_n-E_m .
\end{equation*}
The mass-dependent correction to the kinetic energy affects the $\cA/c^2$ term only beyond the retained order. Substitution into eq.~\eqref{eq:Bornfamily} yields, for $E_m\neq E_n$,
\begin{equation}
\mathcal T^{(1)}_{mn}
=M_{\rm ref}\,\rinv_{mn}\,\tilde\Phi(\bm q).
\label{eq:Borninvariant}
\end{equation}
Thus, the scattering kernel depends on the invariant combination and not on $\mathcal G_{mn}$ separately. For a Newtonian potential we have $\tilde\Phi\propto1/\bm q^2$, so that within the family~\eqref{eq:family} this coefficient multiplies an exchange kernel with the same $1/\bm q^2$ dependence as the potential.

\paragraph{Matching the non-tidal gravitational sector.}
The matching proceeds in three steps. First, the on-shell three-point amplitudes determine the residue of the exchange pole in eq.~\eqref{eq:tresidue}. Second, a regular local continuation separates its curvature sector from the terms supported on the source. Third, eq.~\eqref{eq:Borninvariant} identifies the coefficient in the retained test family. In the non-tidal sector, the graviton has no independent monopole or dipole coefficient between unequal masses. The matching therefore gives
\begin{equation}
\rinv_{mn}=0\qquad(E_m\neq E_n)
\label{eq:rinvzero}
\end{equation}
under the matching assumptions of sections~\ref{sec:assumptions} and~\ref{sec:tchannel}. For transitions with $s_m+s_n<2$, which are forbidden by the selection rule, this also follows directly from the vanishing of the exchange residue. For allowed spins, it is a statement about the specified non-tidal truncation after the curvature operators have been separated. It is not obtained by taking $q^\mu\to0$ with unequal-mass matter states on shell, nor by discarding momentum-dependent curvature terms after the on-shell condition is used.

The diagonal and degenerate blocks are fixed by the universal charge in (S1), while (S2) fixes the universal dipole data. Together with eq.~\eqref{eq:rinvzero}, the matching can be summarized in operator form as
\begin{equation}
\mathcal G_{\rm phys}
\equiv\mathcal G-\frac{\ii}{\hbar c^2}[\Hint,\cA]
=\hat M
\label{eq:Gphysical}
\end{equation}
at the retained order. The operator on the left-hand side is Hermitian and invariant under eq.~\eqref{eq:transformation}. Equation~\eqref{eq:Gphysical} does not assert that $\mathcal G=\hat M$ in every representation. If additional operators contribute at the same order, they must be included in the Born kernel before its invariant coefficients are identified. In that case, $\rinv$ alone need not describe the full transition response. In particular, we do not infer a general matching theorem for arbitrary spin from the curvature identity alone.

\paragraph{Consistency check with the local-source construction.}
The derivation of section~\ref{sec:matching} gives $\mathcal G=\hat M$ and $\cA=0$ for the internal part of the Fermi-Walker Hamiltonian at this order. In the scalar internal sector generated by $\Phi(\bX)D_{\rm int}$, the Hamiltonian in the Newtonian gauge has $\mathcal G=\hat M+\gamma\Sigma/c^2$ and $\cA=\gamma D_{\rm int}$, with the other terms of eq.~\eqref{eq:decomposition} transformed consistently. Since
\begin{equation*}
\Sigma_{mn}=\frac{\ii}{\hbar}(E_m-E_n)(D_{\rm int})_{mn},
\end{equation*}
the two contributions cancel in $\rinv_{mn}$, and both representations give eqs.~\eqref{eq:rinvzero} and~\eqref{eq:Gphysical}. This check relies on the stronger local-source hypothesis and is independent of the matching based on the soft theorems.

The operator $\cA$ has a simple interpretation. Along a trajectory $\bX(t)$ of the center of mass, the internal state evolves with the Hamiltonian $\Hint(\lambda)+\dot\lambda\,\cA(\lambda)$, where $\lambda=\Phi(\bX(t))/c^2$ and $\Hint(\lambda)=\Hint+\lambda c^2\mathcal{G}$ at first order. Under a transformation $U(\lambda)$,
\begin{equation}
\Hint(\lambda)\to U\Hint(\lambda)U^\dagger,\qquad \cA\to U\cA U^\dagger+\ii\hbar\,(\partial_\lambda U)U^\dagger,
\end{equation}
so that $\cA$ is a connection, of the same type as the one that appears in adiabatic evolution~\cite{Berry:1984jv,Wilczek:1984dh,Kolodrubetz:2017ofs}. It cannot be chosen freely, since it is determined by the velocity-dependent terms of the Hamiltonian derived in the given coordinates. Two sets of data are invariant under these transformations. The first is the spectrum $E_n(\lambda)$ of $\Hint(\lambda)$. The second is the non-adiabatic coupling between the instantaneous eigenstates $|n(\lambda)\rangle$,
\begin{equation}
\cN_{mn}(\lambda)=\langle m|\cA|n\rangle-\ii\hbar\,\frac{\langle m|\partial_\lambda\Hint|n\rangle}{E_n-E_m}\qquad(m\neq n),
\label{eq:Nmn}
\end{equation}
which changes only by the phases of the eigenstates (appendix~\ref{app:connection}). At $\lambda=0$ it is related to eq.~\eqref{eq:rinv} by $\cN_{mn}=\ii\hbar M_{\rm ref}c^2\,\rinv_{mn}/(E_m-E_n)$.

In terms of these data the internal content of the QEP in a uniform field can be stated without choosing a field-dependent basis. The total rest energies are redshifted universally. With the above definition of $\Hint(\lambda)$, this means
\begin{equation}
mc^2+E_n(\lambda)=(1+\lambda)\,[mc^2+E_n(0)]+O(\lambda^2),
\label{eq:invQEP}
\end{equation}
with no splitting of degenerate blocks at first order, and $\cN_{mn}=0$ between unequal energies. Equivalently, all transition energies acquire the same factor $1+\lambda$. The common shift $\lambda mc^2$ drops out of clock comparisons but contributes to the universal center-of-mass force. These statements are the invariant form of eq.~\eqref{eq:Gphysical} within the test family. Under the local-source hypothesis, proposition~\ref{prop:mass} independently excludes mass-changing transitions without an adiabatic approximation, while the statement about the energy spectrum refers to a system at rest at each position. Curvature can modify these uniform-field relations through the ordinary tidal response, which is not by itself a violation of the equivalence principle.

This formulation has a direct operational meaning. A resonant pulse whose duration is long compared with the inverse level spacing addresses the eigenstates of $\Hint(\lambda)$ at the position of the atom. The spectrum $E_n(\lambda)$ is measured by clocks and, through the force $-\nabla E_n(\lambda(\bX))$, by the free fall of the levels. The coupling $\cN_{mn}$ is measured by the transitions that the motion through the field induces.

\section{Additional light fields}
\label{sec:fields}

Within the assumptions and the scope of the matching in sections~\ref{sec:softsummary} and~\ref{sec:invariant}, a long-range violation below tidal order cannot be attributed to the single massless spin-two mediator considered here. Additional light fields are one possible origin of such a violation. A failure of the other assumptions would require a separate analysis. In this section, we discuss mediators of lower spin.

\subsection{Scalar field}

Let $\varphi$ be a light scalar field that couples to a local scalar operator $\Theta(x)$ of the matter theory through $\mathcal{L}_{\rm int}=-\varphi\,\Theta$. In a static and slowly varying background $\varphi(\bx)$ the Hamiltonian of a system at rest at $\bX$ acquires the term
\begin{equation}
H'=\int\dd^3x\,\varphi(\bx)\,\Theta(\bx)=\varphi(\bX)\,Q_\Theta+\cdots,\qquad Q_\Theta=\int\dd^3x\,\Theta(\bx),
\label{eq:scalarcoupling}
\end{equation}
where the dots denote terms with gradients of $\varphi$. The amplitude for the emission of a soft scalar has no gauge variation, and the soft theorem imposes no condition on $Q_\Theta$~\cite{Weinberg:1964ew}. The on-shell amplitude for a transition between two spinless levels with the emission of the scalar is likewise unconstrained. The operator $\Theta$ is not conserved, and no Ward identity removes its matrix elements between levels of different mass. Hence, $Q_\Theta$ is an arbitrary Hermitian operator on $\cH_{\rm int}$. In the language of section~\ref{sec:invariant}, with $\lambda$ replaced by $\varphi$, the spectrum is
\begin{equation}
E_n(\varphi)=E_n+\varphi\,(Q_\Theta)_{nn}+\varphi^2\sum_{m\neq n}\frac{|(Q_\Theta)_{mn}|^2}{E_n-E_m}+O(\varphi^3),
\label{eq:scalarspectrum}
\end{equation}
and the invariant off-diagonal parameter is proportional to $(Q_\Theta)_{mn}$. The diagonal elements give accelerations that depend on the level. These are the composition-dependent violations of the weak equivalence principle known from dilaton models~\cite{Damour:2010rp}. The off-diagonal elements give a violation of the quantum type.

For two levels we can write $Q_\Theta=q_0\mathbf{1}+\bm{q}\cdot\bm{\sigma}$. The off-diagonal parameter is proportional to $q_x+\ii q_y$, and the difference between the accelerations of the two levels is proportional to $q_z$ if the small difference in their masses is neglected. These components are independent. A relation between the quantum and the classical violation requires a specific model of $\Theta$.

An important special case is a metric coupling, in which matter couples to $\tilde g_{\mu\nu}=(1+2\alpha\varphi)g_{\mu\nu}$ and $\Theta=-\alpha T^\mu{}_\mu$. This corresponds to the perturbation $h_{00}=-2\alpha\varphi$, $h_{ij}=2\alpha\varphi\,\delta_{ij}$, which is of the form~\eqref{eq:Newtonian} with $\Phi/c^2=\alpha\varphi$ and $\gamma=-1$. Proposition~\ref{prop:unitary} then shows that, for a uniform gradient of $\varphi$, the coupling is unitarily equivalent to a universal one, $\alpha\varphi(\bX)\hat{M}c^2$. The virial part of $Q_\Theta$ does not lead to a violation. This is the statement, at first order in the background, that scalar-tensor theories with a metric coupling respect the weak equivalence principle for test bodies. It is known that this property is protected by Ward identities to all orders in the matter couplings and that it is violated by loops that involve the gravitational scalar~\cite{Armendariz-Picon:2011ydk}. The soft limit of such a scalar is constrained by the dilatation properties of the amplitude~\cite{DiVecchia:2015jaq}.

\subsection{Vector field}

A light vector field couples to a conserved current $J^\mu$. Its charge $Q_V=\int\dd^3x\,J^0$ commutes with $H$, and gauge invariance of the soft factor requires the conservation of the charge in every process~\cite{Weinberg:1964ew}. The charge can differ from level to level only if the levels differ in a conserved quantum number, and it is not proportional to the mass in general. This gives a classical violation of the weak equivalence principle that depends on the composition. The Ward identity implies that the transition charge between levels of different mass vanishes, as noted below eq.~\eqref{eq:RicciForm}, and the argument of section~\ref{sec:selection} gives the same result without a local current. The transition dipole moments are not constrained, and they couple to the field strength. A uniform field strength of the new vector therefore mixes the levels, in the same way as an electric field does in the Stark effect.

\subsection{Minimal order of the off-diagonal coupling}

The three cases are compared in table~\ref{tab:mediators}. In the language of section~\ref{sec:pole}, the on-shell amplitude for a transition between two levels of different mass with the emission of a massless particle of helicity $h$ exists only for $s_n+s_{n'}\ge|h|$~\cite{Conde:2016vxs,Arkani-Hamed:2017jhn}. There is no restriction for a scalar. For a vector the transition is at least of dipole order, and for the graviton it is at least of quadrupole order. In a regular local representation in terms of the field strength, the transition couplings other than the charge involve $|h|$ derivatives of the potential. The corresponding static derivative counting uses the matching assumptions of section~\ref{sec:assumptions}, and the selection rule for the on-shell amplitude alone does not classify all momentum-dependent operators in the Hamiltonian. The order $|h|$ is a lower bound. The conservation of angular momentum and parity can raise the order, since a multipole of order $L$ connects the spins $s_n$ and $s_{n'}$ only for $L\ge|s_n-s_{n'}|$. Furthermore, in some models a scalar couples only through its gradient.

\begin{table}[t]
\centering
\begin{tabular}{llll}
\toprule
Mediator & Monopole on a level & Off-diagonal coupling & Minimal order\\
\midrule
scalar, $h=0$ & arbitrary & $Q_\Theta$ & $\varphi$\\
vector, $|h|=1$ & conserved charge & transition dipoles & $\partial\Phi_V$\\
graviton, $|h|=2$ & $M_n$, universal & mass quadrupole & $\partial\partial\Phi$\\
\bottomrule
\end{tabular}
\caption{Couplings of a system with internal levels to static fields of massless mediators. The last column gives the lowest number of derivatives of the potential at which a coupling between non-degenerate levels can appear. The derivative counting uses the matching assumptions of section~\ref{sec:assumptions}; selection rules can raise this order.}
\label{tab:mediators}
\end{table}

Within this field-strength derivative expansion, an off-diagonal coupling to the value of a potential requires a scalar field, and an off-diagonal coupling to a uniform field strength requires a scalar or a vector field. The order of the multipole gives an upper bound on the helicity of the mediator and does not determine it. For gravity the first off-diagonal coupling is the tidal one. Thus, a non-zero value of $\rinv$ below tidal order, after all operators relevant to the measurement are taken into account, cannot be described by the exchange of a single graviton under our assumptions. Additional light fields are one possible explanation, but this conclusion does not follow from Lorentz invariance alone. Frameworks that violate Lorentz invariance are outside our assumptions~\cite{Hertzberg:2017nzl}.

\section{Observable consequences of off-diagonal couplings}
\label{sec:obs}

Consider a static off-diagonal coupling $W(z)$ between two levels, such as a coupling to an additional light field (section~\ref{sec:fields}). In this section, $\Delta=\hbar\Omega>0$ denotes the energy splitting between the two unperturbed levels. We work in a representation with no velocity-dependent internal coupling, $\cA=0$ in the notation of section~\ref{sec:invariant}. Taking $W$ to be real and measuring energies from the midpoint of the unperturbed levels, we have
\begin{equation}
\Hint(z)=\frac{\Delta}{2}\,\sigma_z+W(z)\,\sigma_x .
\label{eq:twolevel}
\end{equation}
For example, the test Hamiltonian~\eqref{eq:Htest} gives $W(z)=rM_{\rm ref}\Phi(z)$ for real $r$, with the common reference mass of section~\ref{sec:invariant}. In this representation $r$ coincides with the invariant parameter in eq.~\eqref{eq:rinv}.

The energy eigenvalues are
\begin{equation}
E_\pm(z)=\pm\frac12\sqrt{\Delta^2+4W(z)^2}
\simeq\pm\left(\frac{\Delta}{2}+\frac{W(z)^2}{\Delta}\right),
\label{eq:dressed}
\end{equation}
where the expansion holds for $|W|\ll\Delta$. An atom that adiabatically follows one of these levels experiences a force due to the spatial variation of the level shift. The same level repulsion shifts the transition frequency from its unperturbed value $\nu_0=\Delta/(2\pi\hbar)$. To second order in the coupling,
\begin{equation}
F_\pm(z)=-\frac{\dd E_\pm}{\dd z}\simeq\mp\frac{2W(z)W'(z)}{\Delta},
\qquad
\frac{\nu(z)-\nu_0}{\nu_0}\simeq\frac{2W(z)^2}{\Delta^2}.
\label{eq:clock}
\end{equation}
Free-fall measurements on individual energy eigenstates are therefore sensitive to the off-diagonal coupling at second order. Independent interactions that are quadratic in the external field could modify these predictions of the model~\eqref{eq:twolevel}.

For an atom interferometer, the first-order signal depends on the internal evolution along the two paths $z_a(t)$ and $z_b(t)$. The matrix element for the transition from the lower to the upper level carries a phase $e^{\ii\Omega t}$ in the interaction picture. For paths that split at $t=0$ and recombine at $2T$, the difference between the first-order transition amplitudes is
\begin{align}
\epsilon_a-\epsilon_b
&=-\frac{\ii}{\hbar}\int_0^{2T}\dd t\,\big[W(z_a(t))-W(z_b(t))\big]e^{\ii\Omega t}
\nonumber\\
&\simeq-\frac{\ii W'_0}{\hbar}\int_0^{2T}\dd t\,\delta z(t)\,e^{\ii\Omega t}.
\label{eq:firstorder}
\end{align}
Here $\delta z=z_a-z_b$ and the second line assumes that the gradient $W'_0$ is nearly constant over both paths. The overlap of the internal states on the two paths contributes to the fringe phase~\cite{Roura:2018cfg}. For an initial coherent superposition, its first-order phase shift depends on $\epsilon_a-\epsilon_b$ and on the relative phase of the two levels.

\begin{figure}[htbp]
\centering
\includegraphics[width=0.62\textwidth]{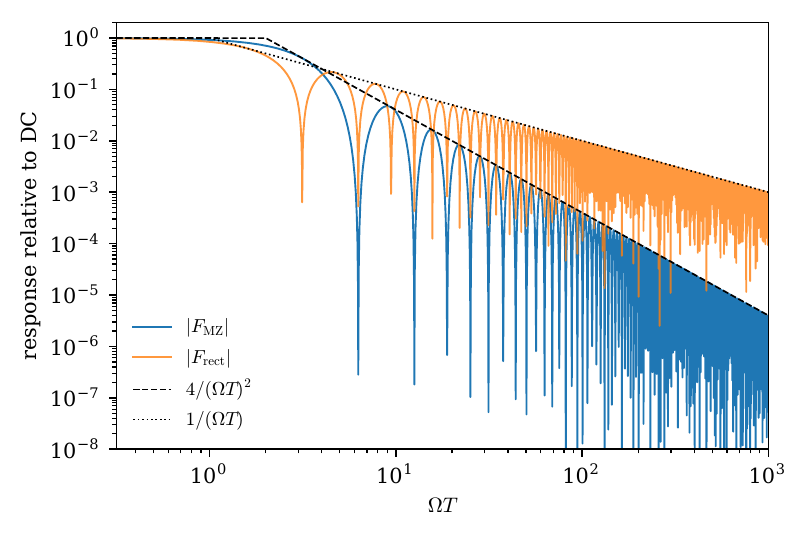}
\caption{Response of a Mach-Zehnder atom interferometer to a static off-diagonal coupling, normalized to its zero-frequency value. The blue curve shows $|F_{\rm MZ}|$ from eq.~\eqref{eq:FMZ}. The orange curve shows $|F_{\rm rect}|=|\sin(\Omega T)/(\Omega T)|$ for uniform weighting over the same interval $0\le t\le2T$. The dashed and dotted lines give the respective upper envelopes $4/(\Omega T)^2$ and $1/(\Omega T)$, each capped at unity.}
\label{fig:sensitivity}
\end{figure}

For an ideal Mach-Zehnder sequence with instantaneous pulses at $0$, $T$ and $2T$ that preserve the internal levels, the path separation is triangular, $\delta z(t)=v_{\rm rec}\min(t,2T-t)$, where $v_{\rm rec}$ is the recoil velocity. The response normalized to its zero-frequency value is~\cite{Cheinet:2005dk}
\begin{equation}
F_{\rm MZ}(\Omega)
=\frac{1}{v_{\rm rec}T^2}\int_0^{2T}\dd t\,\delta z(t)\,e^{\ii\Omega t}
=e^{\ii\Omega T}\left[\frac{\sin(\Omega T/2)}{\Omega T/2}\right]^2 .
\label{eq:FMZ}
\end{equation}
Figure~\ref{fig:sensitivity} shows the suppression of $|F_{\rm MZ}|$ with increasing $\Omega T$. Its magnitude is bounded by $\min\{1,4/(\Omega T)^2\}$. Thus, when $\Omega T\gg1$, the internal evolution averages out most of the first-order response to a static coupling, even for a coherent superposition.

Rosi et al.~\cite{Rosi:2017ieh} inferred a bound on the off-diagonal parameter from the shot-to-shot scatter of the interferometric phase. Onofrio, Smith and Viola~\cite{Onofrio:2025ixt} showed that averaging over an uncontrolled initial relative phase removes the linear term from the mean acceleration, while the acceleration variance retains a dependence on $|r|^2$. The suppression in eq.~\eqref{eq:FMZ} occurs during a single pulse sequence, even when the initial phase is fixed.

For nearly degenerate levels with $\Omega T\lesssim1$, the relative phase changes little during the sequence. A coupling modulated at the frequency $\Omega$ can also produce a first-order response by resonantly driving the transition. A different linear effect arises in the energy spectrum if an existing interaction mixes the two levels and interferes with $W$. To compare the sensitivities of measurements with eigenstates and with superpositions in a particular experiment, one must take into account the pulse sequence, the state preparation and the readout.

\section{Conclusions}
\label{sec:conclusions}

In this paper, we have investigated how much of the quantum equivalence principle follows from the on-shell consistency of a massless spin-two particle. The leading soft theorem fixes the gravitational charge on each stable level and each degenerate multiplet, and the subleading theorem fixes the coupling to the total Lorentz generator. However, these theorems do not determine the transitions between levels of different mass, for which there is no soft pole. Moreover, the transition vertex at zero momentum transfer is not an observable of the $S$-matrix, because it can be moved into contact terms.

The invariant information about these transitions is carried by the mass-changing on-shell amplitudes, in which the system changes its level by emitting or absorbing a graviton. They obey the selection rule $s_n+s_{n'}\ge2$ and admit a curvature representation. Under the regularity and power-counting assumptions of section~\ref{sec:assumptions}, we have matched the long-range graviton exchange that they determine to the test Hamiltonian~\eqref{eq:family}. In this family, an off-diagonal element of the gravitational mass operator alone is not invariant under field-dependent changes of the internal basis. The transition kernel~\eqref{eq:Borninvariant} depends only on $\rinv$, the invariant combination of the potential and momentum-dependent couplings. The matching gives $\rinv_{mn}=0$ in the non-tidal sector and, together with the diagonal and degenerate blocks, $\mathcal G_{\rm phys}=\hat M$.

Independently of the soft theorems, we have assumed that gravity couples to a local conserved stress tensor whose charges generate Poincar\'e transformations. Below tidal order, the laboratory Hamiltonian is then $\Hlab=H+\ba\cdot\bK-\bom\cdot\bJ$, which commutes with the mass operator. Hence, it does not induce transitions between levels of different mass, and the same operator $\hat M=m\mathbf1+\Hint/c^2$ appears as the rest, inertial and gravitational mass in the Fermi-Walker frame. The virial terms in the Newtonian gauge are removed by the unitary transformation generated by the dilatation and special conformal moments, and are therefore a consequence of the choice of coordinates. In invariant terms, the total rest energies are redshifted universally and the covariant non-adiabatic coupling vanishes, in agreement with the matching.

Curvature couplings can induce transitions between levels that are allowed by angular momentum. These tidal transitions are ordinary finite-size effects and are compatible with the equivalence principle in a uniform field. Additional light scalar or vector fields can produce off-diagonal couplings below tidal order. In a two-level model, a static off-diagonal coupling shifts the transition frequency and produces a level-dependent force at second order, whereas the first-order response of an ideal Mach-Zehnder interferometer is suppressed by the rapid internal evolution.

Our analysis has several limitations. The matching is at tree level and at first order in the external field. It assumes a regular local derivative expansion and the closure of the test family, which do not follow from the curvature identity~\eqref{eq:RiemannForm} alone. In four dimensions, massless loops produce non-analytic long-range terms~\cite{Donoghue:1994dn,Bjerrum-Bohr:2014zsa} and infrared corrections to the soft expansion. We have not classified all momentum-dependent transition operators for arbitrary spin, considered a simultaneous soft and zero-gap limit, or treated unstable levels. The local-source hypothesis is not derived from the $S$-matrix. Quantum sources, quantum reference frames~\cite{Giacomini:2020ahk,Hardy:2019cef} and infrared-dressed states require a separate analysis.

\acknowledgments
\subsection*{Use of generative AI}

During the preparation of this work, the author used Claude (Anthropic) to assist with drafting and revising the text, language editing, and improving the readability of the manuscript. Claude was also used to assist with the derivations, numerical calculations, and preparation of figure~\ref{fig:sensitivity}. It was further used for proofreading, including checks for typographical errors and checks of the consistency of the notation, equations, and bibliographic information. ChatGPT (OpenAI) was also used to assist with manuscript revisions, including clarifying the notation and refining the exposition. The author reviewed and verified all AI-assisted output and takes full responsibility for the content of this article.

\appendix

\section{Conventions and linearized curvature}
\label{app:conventions}

We use $\eta_{\mu\nu}={\rm diag}(-1,1,1,1)$ and $g_{\mu\nu}=\eta_{\mu\nu}+h_{\mu\nu}$. In sections~\ref{sec:soft}--\ref{sec:source} the canonically normalized graviton field is $h_{\mu\nu}/\kappa$ with $\kappa^2=32\pi G_{\rm N}$. The linearized Riemann tensor is
\begin{equation}
R_{\mu\nu\rho\sigma}=\frac12\big(\partial_\nu\partial_\rho h_{\mu\sigma}+\partial_\mu\partial_\sigma h_{\nu\rho}-\partial_\mu\partial_\rho h_{\nu\sigma}-\partial_\nu\partial_\sigma h_{\mu\rho}\big),
\end{equation}
and for the plane wave $h_{\mu\nu}=\varepsilon_{\mu\nu}e^{\ii q\cdot x}$ it becomes
\begin{equation}
R_{\mu\nu\rho\sigma}[\varepsilon]=-\frac12\big(q_\nu q_\rho\varepsilon_{\mu\sigma}+q_\mu q_\sigma\varepsilon_{\nu\rho}-q_\mu q_\rho\varepsilon_{\nu\sigma}-q_\nu q_\sigma\varepsilon_{\mu\rho}\big),
\end{equation}
where the exponential is omitted. The Ricci tensor $R_{\mu\nu}=\eta^{\rho\sigma}R_{\rho\mu\sigma\nu}$ and the Ricci scalar are given in eq.~\eqref{eq:Ricci}. Contracting $R_{\mu\nu\rho\sigma}[\varepsilon]$ with $\ell^\nu\ell^\rho$ for $\ell\cdot q=1$ gives eq.~\eqref{eq:lemma}. To verify eq.~\eqref{eq:RicciForm}, we note that
\begin{equation}
R_{\mu\nu}P^\mu P^\nu=\frac12\big[q^2\,\varepsilon_{\mu\nu}P^\mu P^\nu+\Delta^2\,\varepsilon^\rho{}_\rho+2\Delta\,\varepsilon_{\mu\nu}q^\mu P^\nu\big],
\end{equation}
where we used $P\cdot q=-\Delta$, and compare this result with the contraction of eq.~\eqref{eq:general} after the substitution~\eqref{eq:solution}.

\section{Poincar\'e generators of a massive level}
\label{app:poincare}

In the momentum representation, with $\bX=\ii\hbar\,\partial/\partial\bP$, the generators~\eqref{eq:NW} act on wave functions with $2s+1$ components. The commutators $[K_i,H]=\ii\hbar P_i$ and $[K_i,P_j]=\ii\hbar\delta_{ij}H/c^2$ follow from the first term of $\bK$, since the second term commutes with $\bP$ and $H$. The commutator of two boosts receives contributions from both terms,
\begin{equation}
[K_i,K_j]=-\frac{\ii\hbar}{c^2}\,\epsilon_{ijk}\,(\bX\times\bP)_k-\frac{\ii\hbar}{c^2}\,\epsilon_{ijk}\,S_k,
\end{equation}
where the spin term arises from the commutator of $\bX$ with $(\bS\times\bP)/(H+M_nc^2)$ and from the commutator of the spin operators. The relative sign of the two terms in $\bK$ is fixed by this relation. We have verified all the commutators~\eqref{eq:algebra} for spin $\tfrac12$ by acting on arbitrary two-component functions of the momentum.

If the boost is displaced by a multiple of the spin, $\bK'=\bK+c_{\rm d}\bS$, the first two relations of eq.~\eqref{eq:algebra} are unchanged, and at $\bP=0$
\begin{equation}
[K'_i,K'_j]=-\frac{\ii\hbar}{c^2}\,\epsilon_{ijk}J_k+\ii\hbar\,c_{\rm d}^2\,\epsilon_{ijk}S_k .
\end{equation}
The algebra is satisfied only for $c_{\rm d}=0$. This is the statement that the center of energy of a level cannot be displaced along its spin, which corresponds to the case (b) of section~\ref{sec:vertexq}.

The expansion~\eqref{eq:Hexp} on a level of mass $M_n$ follows from $H=M_nc^2+\bP^2/2M_n-\bP^4/8M_n^3c^2+\cdots$ and
\begin{equation}
\ba\cdot\bK=M_n\,\ba\cdot\bX+\frac{1}{4M_nc^2}\{\ba\cdot\bX,\bP^2\}-\frac{\ba\cdot(\bS\times\bP)}{2M_nc^2}+O(c^{-4}),
\end{equation}
together with $\ba\cdot(\bS\times\bP)=-\bS\cdot(\ba\times\bP)$.

\section{Dilatation and special conformal moments}
\label{app:conformal}

With $x^0=ct$ the conservation law reads $\partial_t(T^{0i}/c)=-\partial_jT^{ji}$. For the densities of eq.~\eqref{eq:DC} we find
\begin{align}
\partial_t\Big(\frac{x^iT^{0i}}{c}\Big)&=-\partial_j\big(x^iT^{ji}\big)+T^{ii},\\
\partial_t\Big(\frac{x^kx^iT^{0i}-\tfrac12\bx^2T^{0k}}{c}\Big)&=-\partial_j\Big(x^kx^iT^{ji}-\frac12\bx^2T^{jk}\Big)+x^k\,T^{ii},
\end{align}
where the symmetry of $T^{ij}$ was used in the second line. Integrating over space and using $\dd O/\dd t=(\ii/\hbar)[H,O]$ for the operators $D$ and $C^k$, which have no explicit time dependence, gives eq.~\eqref{eq:DCidentities}.

For a system of particles with positions $\bx_a$ and momenta $\bm{p}_a$, the moments are $D=\sum_a\tfrac12\{\bx_a,\bm{p}_a\}$ and $C^k=\sum_a(x_a^k\,\bx_a\cdot\bm{p}_a-\tfrac12\bx_a^2p_a^k)$ with symmetrized products, and $U_\gamma$ generates the change of coordinates given below eq.~\eqref{eq:Ugamma}. Writing $\bx_a=\bX+\bm{\xi}_a$, one obtains the decomposition~\eqref{eq:decomposition}, with $D_{\rm int}=\sum_a\tfrac12\{\bm{\xi}_a,\bm{\pi}_a\}$ in terms of the relative momenta $\bm{\pi}_a$.

As a check of proposition~\ref{prop:unitary}, consider a free relativistic particle of mass $M$ in the metric~\eqref{eq:Newtonian}. Its Hamiltonian is
\begin{equation}
\HN=\sqrt{1+\frac{2\Phi}{c^2}}\,\sqrt{M^2c^4+\frac{c^2\bm{p}^2}{1-2\gamma\Phi/c^2}}
=E+\frac{\Phi}{c^2}\Big(E+\gamma\,\frac{c^2\bm{p}^2}{E}\Big)+O(\Phi^2),
\end{equation}
with $E=(M^2c^4+c^2\bm{p}^2)^{1/2}$, in agreement with eq.~\eqref{eq:HN}. For $\Phi=\Phi_0+\ba\cdot\bx$ the classical moments satisfy $\{\Phi_0D+\ba\cdot\bC,E\}=\Phi\,c^2\bm{p}^2/E$ and $\Phi_0E/c^2+\ba\cdot\bK=\Phi E/c^2$, where $\{\,,\}$ is the Poisson bracket and $\bK=\bx E/c^2$. Hence, the term proportional to $\gamma$ is generated by the canonical transformation with the generator $\gamma(\Phi_0D+\ba\cdot\bC)/c^2$.

\section{Covariance of the non-adiabatic coupling}
\label{app:connection}

Let the internal state evolve with $\Hint(\lambda)+\dot\lambda\,\cA(\lambda)$, and expand it in the instantaneous eigenstates, $|\psi\rangle=\sum_nc_n\,e^{-\ii\theta_n}|n(\lambda)\rangle$ with $\hbar\dot\theta_n=E_n(\lambda)$. The coefficients satisfy
\begin{equation}
\ii\hbar\,\dot c_m=\dot\lambda\sum_n\Big[\langle m|\cA|n\rangle-\ii\hbar\,\langle m|\partial_\lambda n\rangle\Big]e^{\ii(\theta_m-\theta_n)}\,c_n .
\end{equation}
For $m\neq n$ the bracket is the coupling $\cN_{mn}$ of eq.~\eqref{eq:Nmn}, because the eigenstates satisfy $\langle m|\partial_\lambda n\rangle=\langle m|\partial_\lambda\Hint|n\rangle/(E_n-E_m)$. Under a transformation $U(\lambda)$ the eigenstates become $U|n\rangle$, and the connection acquires the additional term $\ii\hbar(\partial_\lambda U)U^\dagger$. The two terms that contain $\partial_\lambda U$ cancel,
\begin{equation}
\langle m|U^\dagger\big[\ii\hbar(\partial_\lambda U)U^\dagger\big]U|n\rangle-\ii\hbar\,\langle m|U^\dagger(\partial_\lambda U)|n\rangle=0,
\end{equation}
so that $\cN_{mn}$ is unchanged up to the phases of the eigenstates. The populations of the instantaneous eigenstates after an arbitrary variation $\lambda(t)$ are therefore the same in all representations. We have confirmed this by a numerical integration with random Hermitian matrices.

At first order in $\dot\lambda$ the amplitude for a transition from the level $n$ to the level $m$ is
\begin{equation}
c_m=-\frac{\ii}{\hbar}\int\dd t\,\dot\lambda\,\cN_{mn}\,e^{\ii(\theta_m-\theta_n)} .
\end{equation}
For the two-level model~\eqref{eq:twolevel} with $\cA=0$ and $|W|\ll\Delta$, the coupling per unit height is $\cN_{21}\,\dd\lambda/\dd z=\ii\hbar W'/\Delta$, and the amplitude becomes $(W'/\Delta)\int\dd t\,\dot z\,e^{\ii\Omega t}$. For the difference between the amplitudes along two paths with common initial and final points, integration by parts gives eq.~\eqref{eq:firstorder}.

\section{A model with two levels}
\label{app:toy}

In this appendix we illustrate the statements of sections~\ref{sec:softnondeg} and~\ref{sec:low} in an explicit model. We set $\hbar=c=1$. The two levels are described by real scalar fields $\phi_1$ and $\phi_2$ of masses $M_1\neq M_2$, which interact with each other and with a third scalar field $\chi$ of mass $\mu$,
\begin{equation}
\mathcal{L}_{\rm int}=-\frac{\lambda}{3!}\,\chi^3-\sum_{a=1,2}\frac{g_a}{2}\,\phi_a\chi^2-g_{12}\,\phi_1\phi_2\chi .
\label{eq:toymodel}
\end{equation}
All fields are minimally coupled to the metric $\eta_{\mu\nu}+\kappa h_{\mu\nu}$. We consider the amplitude with three particles $\chi$ of outgoing momenta $k_1$, $k_2$, $k_3$ and one particle in the level $a$ of momentum $k_4$. At tree level,
\begin{equation}
\cM_4^{(a)}=\sum_{1\le i<j\le3}\bigg[\frac{\lambda\,g_a}{(k_i+k_j)^2+\mu^2}+\frac{g_b\,g_{12}}{(k_i+k_j)^2+M_b^2}\bigg],\qquad b\neq a.
\label{eq:toyM4}
\end{equation}
In the second term the other level is exchanged, so that the amplitude of each level depends on the couplings of both. For a transverse and traceless polarization the graviton couples to a line of momentum $P$ through $\kappa\,\varepsilon_{\mu\nu}P^\mu P^\nu$, and the vertices of eq.~\eqref{eq:toymodel} give no contact terms. A diagram contributing to eq.~\eqref{eq:toyM4} with coupling $c$, internal mass $m$ and internal momentum $P=k_i+k_j$ gives
\begin{equation}
\kappa\,c\,\bigg[\sum_{x=i,j}\frac{\varepsilon_{\mu\nu}k_x^\mu k_x^\nu}{2k_x\cdot q}\,\frac{1}{D(P+q)}+\sum_{x\neq i,j}\frac{\varepsilon_{\mu\nu}k_x^\mu k_x^\nu}{2k_x\cdot q}\,\frac{1}{D(P)}+\frac{\varepsilon_{\mu\nu}P^\mu P^\nu}{D(P)D(P+q)}\bigg],
\label{eq:toyM5}
\end{equation}
where $D(P)=P^2+m^2$. The sum of these terms is invariant under the shift of the polarization because momentum conservation gives $\sum_xk_x+q=0$. In these variables the transition vertices vanish, $V^{\mu\nu}_{12}=0$.

We have evaluated $\cM_5^{(a)}$ numerically and compared it with eq.~\eqref{eq:subleading}, where for scalars
\begin{equation}
S^{(1)}\cM_4=\frac{\kappa}{2}\sum_x\bigg[\frac{\varepsilon_{\mu\nu}k_x^\mu k_x^\nu}{k_x\cdot q}\,q_\rho\frac{\partial}{\partial k_{x\rho}}-\varepsilon_{\mu\nu}k_x^\mu\frac{\partial}{\partial k_{x\nu}}\bigg]\cM_4,
\end{equation}
and $\cM_4$ is the function~\eqref{eq:toyM4} of the momenta. The result does not depend on how this function is continued away from $\sum_xk_x=0$. Table~\ref{tab:toy} shows the result for $\mu=1$, $M_1=1.9$, $M_2=2.6$, $\lambda=0.7$, $g_1=1.3$, $g_2=-0.9$ and $g_{12}=0.5$, for fixed directions of the momenta. After the subtraction of the leading term the remainder is independent of $\omega$, and after the subtraction of the subleading term, which is computed from the amplitude $\cM_4^{(a)}$ of the same level, it is proportional to $\omega$. The amplitude $\cM_4^{(b)}$ of the other level does not appear.

\begin{table}[t]
\centering
\begin{tabular}{lllll}
\toprule
Level & $\omega$ & $|\cM_5|$ & $|\cM_5-S^{(0)}\cM_4|$ & $|\cM_5-(S^{(0)}+S^{(1)})\cM_4|$\\
\midrule
$a=1$ & $1.00\times10^{-2}$ & $20.1$ & $5.25\times10^{-2}$ & $3.03\times10^{-4}$\\
 & $5.00\times10^{-3}$ & $40.2$ & $5.24\times10^{-2}$ & $1.52\times10^{-4}$\\
 & $2.50\times10^{-3}$ & $80.3$ & $5.24\times10^{-2}$ & $7.58\times10^{-5}$\\
 & $1.25\times10^{-3}$ & $161$ & $5.24\times10^{-2}$ & $3.79\times10^{-5}$\\
\midrule
$a=2$ & $1.00\times10^{-2}$ & $4.97$ & $2.40\times10^{-2}$ & $1.29\times10^{-4}$\\
 & $5.00\times10^{-3}$ & $9.94$ & $2.40\times10^{-2}$ & $6.48\times10^{-5}$\\
 & $2.50\times10^{-3}$ & $19.9$ & $2.40\times10^{-2}$ & $3.25\times10^{-5}$\\
 & $1.25\times10^{-3}$ & $39.8$ & $2.40\times10^{-2}$ & $1.62\times10^{-5}$\\
\bottomrule
\end{tabular}
\caption{Soft expansion of the amplitude with one graviton in the model~\eqref{eq:toymodel}, in units of $\kappa$ and with $\mu=1$. The amplitudes are evaluated for a graviton of positive helicity and energy $\omega$ and for fixed directions of the other momenta.}
\label{tab:toy}
\end{table}

We now perform the redefinition~\eqref{eq:redefinition} with $n=2$ and $n'=1$, $\phi_1\to\phi_1+(b\kappa/2)h^{\mu\nu}\partial_\mu\partial_\nu\phi_2$. At first order in $\kappa$, it adds the following terms to the Lagrangian:
\begin{equation}
\frac{b\kappa}{2}\,h^{\mu\nu}\partial_\mu\partial_\nu\phi_2\,\Big[\big(\Box-M_1^2\big)\phi_1-\frac{g_1}{2}\chi^2-g_{12}\,\phi_2\chi\Big].
\end{equation}
The first term is a transition vertex between the two levels. On an external line in level $2$, it is multiplied by the propagator for level $1$. The resulting contribution to $\cM^{\mu\nu}$ is $b\,k_4^\mu k_4^\nu\,\cM_4^{(1)}$, which corresponds to the vertex of eq.~\eqref{eq:redefvertex}. On an external line in level $1$, it vanishes. The remaining terms are contact vertices, and they contribute $-b\,k_4^\mu k_4^\nu\,\cM_4^{(1)}$. For the parameters of table~\ref{tab:toy} and $b$ of order one, each of the two contributions is comparable in size to the term of order $\omega^0$ of the amplitude, and their sum vanishes. The amplitude and its soft expansion are unchanged, although the Lagrangian now contains a coupling of the graviton that connects the two levels at zero momentum transfer.

In contrast, a coupling $\kappa\,g_T\,h_{\mu\nu}\partial^\mu\phi_1\partial^\nu\phi_2$ that is added to the model is not of this form. Its contribution to the amplitude is not invariant under the shift of the polarization, in accordance with eq.~\eqref{eq:deltaA3}. The operator $\partial^\mu\phi_1\partial^\nu\phi_2$ is not conserved for $M_1\neq M_2$, and by the argument of section~\ref{sec:selection} no completion of this coupling is consistent with the Lorentz invariance of the amplitudes.

\bibliographystyle{JHEP}
\bibliography{references}

@article{Will:2014kxa,
    author = "Will, Clifford M.",
    title = "{The Confrontation between General Relativity and Experiment}",
    eprint = "1403.7377",
    archivePrefix = "arXiv",
    primaryClass = "gr-qc",
    doi = "10.12942/lrr-2014-4",
    journal = "Living Rev. Rel.",
    volume = "17",
    pages = "4",
    year = "2014"
}

@article{Zych:2015fka,
    author = "Zych, Magdalena and Brukner, {\v{C}}aslav",
    title = "{Quantum formulation of the Einstein Equivalence Principle}",
    eprint = "1502.00971",
    archivePrefix = "arXiv",
    primaryClass = "gr-qc",
    doi = "10.1038/s41567-018-0197-6",
    journal = "Nature Phys.",
    volume = "14",
    number = "10",
    pages = "1027--1031",
    year = "2018"
}

@article{Rosi:2017ieh,
    author = "Rosi, G. and D'Amico, G. and Cacciapuoti, L. and Sorrentino, F. and Prevedelli, M. and Zych, M. and Brukner, {\v{C}}. and Tino, G. M.",
    title = "{Quantum test of the equivalence principle for atoms in coherent superposition of internal energy states}",
    eprint = "1704.02296",
    archivePrefix = "arXiv",
    primaryClass = "physics.atom-ph",
    doi = "10.1038/ncomms15529",
    journal = "Nature Commun.",
    volume = "8",
    pages = "15529",
    year = "2017"
}

@article{Orlando:2015jdi,
    author = "Orlando, Patrick J. and Mann, Robert B. and Modi, Kavan and Pollock, Felix A.",
    title = "{A test of the equivalence principle(s) for quantum superpositions}",
    eprint = "1511.02943",
    archivePrefix = "arXiv",
    primaryClass = "quant-ph",
    doi = "10.1088/0264-9381/33/19/19LT01",
    journal = "Class. Quant. Grav.",
    volume = "33",
    number = "19",
    pages = "19LT01",
    year = "2016"
}

@article{Das:2023cfu,
    author = "Das, Saurya and Fridman, Mitja and Lambiase, Gaetano",
    title = "{General formalism of the quantum equivalence principle}",
    eprint = "2307.09632",
    archivePrefix = "arXiv",
    primaryClass = "gr-qc",
    doi = "10.1038/s42005-023-01306-w",
    journal = "Commun. Phys.",
    volume = "6",
    number = "1",
    pages = "198",
    year = "2023"
}

@article{Onofrio:2025ixt,
    author = "Onofrio, Roberto and Smith, Alexander R. H. and Viola, Lorenza",
    title = "{Testing the weak equivalence principle for nonclassical matter with torsion balances}",
    eprint = "2512.06333",
    archivePrefix = "arXiv",
    primaryClass = "quant-ph",
    doi = "10.1103/r2ts-fsjj",
    journal = "Phys. Rev. D",
    volume = "112",
    number = "12",
    pages = "124014",
    year = "2025"
}

@article{Weinberg:1964ew,
    author = "Weinberg, Steven",
    title = "{Photons and Gravitons in  $S$-Matrix Theory: Derivation of Charge Conservation and Equality of Gravitational and Inertial Mass}",
    doi = "10.1103/PhysRev.135.B1049",
    journal = "Phys. Rev.",
    volume = "135",
    pages = "B1049--B1056",
    year = "1964"
}

@article{Weinberg:1965nx,
    author = "Weinberg, Steven",
    title = "{Infrared photons and gravitons}",
    doi = "10.1103/PhysRev.140.B516",
    journal = "Phys. Rev.",
    volume = "140",
    pages = "B516--B524",
    year = "1965"
}

@article{Cachazo:2014fwa,
    author = "Cachazo, Freddy and Strominger, Andrew",
    title = "{Evidence for a New Soft Graviton Theorem}",
    eprint = "1404.4091",
    archivePrefix = "arXiv",
    primaryClass = "hep-th",
    month = "4",
    year = "2014"
}

@article{Bern:2014vva,
    author = "Bern, Zvi and Davies, Scott and Di Vecchia, Paolo and Nohle, Josh",
    title = "{Low-Energy Behavior of Gluons and Gravitons from Gauge Invariance}",
    eprint = "1406.6987",
    archivePrefix = "arXiv",
    primaryClass = "hep-th",
    reportNumber = "UCLA-14-TEP-104, NORDITA-2014-78",
    doi = "10.1103/PhysRevD.90.084035",
    journal = "Phys. Rev. D",
    volume = "90",
    number = "8",
    pages = "084035",
    year = "2014"
}

@article{Sen:2017xjn,
    author = "Sen, Ashoke",
    title = "{Soft Theorems in Superstring Theory}",
    eprint = "1702.03934",
    archivePrefix = "arXiv",
    primaryClass = "hep-th",
    doi = "10.1007/JHEP06(2017)113",
    journal = "JHEP",
    volume = "06",
    pages = "113",
    year = "2017"
}

@article{Sen:2017nim,
    author = "Sen, Ashoke",
    title = "{Subleading Soft Graviton Theorem for Loop Amplitudes}",
    eprint = "1703.00024",
    archivePrefix = "arXiv",
    primaryClass = "hep-th",
    doi = "10.1007/JHEP11(2017)123",
    journal = "JHEP",
    volume = "11",
    pages = "123",
    year = "2017"
}

@article{Laddha:2017ygw,
    author = "Laddha, Alok and Sen, Ashoke",
    title = "{Sub-subleading Soft Graviton Theorem in Generic Theories of Quantum Gravity}",
    eprint = "1706.00759",
    archivePrefix = "arXiv",
    primaryClass = "hep-th",
    doi = "10.1007/JHEP10(2017)065",
    journal = "JHEP",
    volume = "10",
    pages = "065",
    year = "2017"
}

@article{Falkowski:2020aso,
    author = "Falkowski, Adam and Machado, Camila S.",
    title = "{Soft Matters, or the Recursions with Massive Spinors}",
    eprint = "2005.08981",
    archivePrefix = "arXiv",
    primaryClass = "hep-th",
    doi = "10.1007/JHEP05(2021)238",
    journal = "JHEP",
    volume = "05",
    pages = "238",
    year = "2021"
}

@article{Conde:2016vxs,
    author = "Conde, Eduardo and Marzolla, Andrea",
    title = "{Lorentz Constraints on Massive Three-Point Amplitudes}",
    eprint = "1601.08113",
    archivePrefix = "arXiv",
    primaryClass = "hep-th",
    doi = "10.1007/JHEP09(2016)041",
    journal = "JHEP",
    volume = "09",
    pages = "041",
    year = "2016"
}

@article{Arkani-Hamed:2017jhn,
    author = "Arkani-Hamed, Nima and Huang, Tzu-Chen and Huang, Yu-tin",
    title = "{Scattering amplitudes for all masses and spins}",
    eprint = "1709.04891",
    archivePrefix = "arXiv",
    primaryClass = "hep-th",
    reportNumber = "NCTS-TH/1714, NCTS-TH-1714",
    doi = "10.1007/JHEP11(2021)070",
    journal = "JHEP",
    volume = "11",
    pages = "070",
    year = "2021"
}

@article{Kobzarev:1962wt,
    author = "Kobzarev, I. Yu. and Okun', L. B.",
    title = "{Gravitational Interaction of Fermions}",
    journal = "Zh. Eksp. Teor. Fiz.",
    volume = "43",
    pages = "1904--1909",
    year = "1962"
}

@article{Pagels:1966zza,
    author = "Pagels, Heinz",
    title = "{Energy-Momentum Structure Form Factors of Particles}",
    doi = "10.1103/PhysRev.144.1250",
    journal = "Phys. Rev.",
    volume = "144",
    pages = "1250--1260",
    year = "1966"
}

@article{Teryaev:1999su,
    author = "Teryaev, O. V.",
    title = "{Spin structure of nucleon and equivalence principle}",
    eprint = "hep-ph/9904376",
    archivePrefix = "arXiv",
    month = "4",
    year = "1999"
}

@article{Teryaev:2016edw,
    author = "Teryaev, O. V.",
    title = "{Gravitational form factors and nucleon spin structure}",
    doi = "10.1007/s11467-016-0573-6",
    journal = "Front. Phys. (Beijing)",
    volume = "11",
    number = "5",
    pages = "111207",
    year = "2016"
}

@article{Silenko:2006er,
    author = "Silenko, Alexander J. and Teryaev, Oleg V.",
    title = "{Equivalence principle and experimental tests of gravitational spin effects}",
    eprint = "gr-qc/0612103",
    archivePrefix = "arXiv",
    doi = "10.1103/PhysRevD.76.061101",
    journal = "Phys. Rev. D",
    volume = "76",
    pages = "061101",
    year = "2007"
}

@article{Cotogno:2019xcl,
    author = "Cotogno, Sabrina and Lorc{\'e}, C{\'e}dric and Lowdon, Peter",
    title = "{Poincar{\'e} constraints on the gravitational form factors for massive states with arbitrary spin}",
    eprint = "1905.11969",
    archivePrefix = "arXiv",
    primaryClass = "hep-th",
    doi = "10.1103/PhysRevD.100.045003",
    journal = "Phys. Rev. D",
    volume = "100",
    number = "4",
    pages = "045003",
    year = "2019"
}

@article{Lorce:2019sbq,
    author = "Lorc{\'e}, C{\'e}dric and Lowdon, Peter",
    title = "{Universality of the Poincar{\'e} gravitational form factor constraints}",
    eprint = "1908.02567",
    archivePrefix = "arXiv",
    primaryClass = "hep-th",
    doi = "10.1140/epjc/s10052-020-7779-z",
    journal = "Eur. Phys. J. C",
    volume = "80",
    number = "3",
    pages = "207",
    year = "2020"
}

@article{Kim:2022bwn,
    author = "Kim, June-Young",
    title = "{Parametrization of transition energy-momentum tensor form factors}",
    eprint = "2206.10202",
    archivePrefix = "arXiv",
    primaryClass = "hep-ph",
    doi = "10.1016/j.physletb.2022.137442",
    journal = "Phys. Lett. B",
    volume = "834",
    pages = "137442",
    year = "2022"
}

@article{Ozdem:2022zig,
    author = {{\"O}zdem, U. and Azizi, K.},
    title = "{Gravitational transition form factors of N {\textrightarrow} {\ensuremath{\Delta}} via QCD light-cone sum rules}",
    eprint = "2212.07290",
    archivePrefix = "arXiv",
    primaryClass = "hep-ph",
    doi = "10.1007/JHEP03(2023)048",
    journal = "JHEP",
    volume = "03",
    pages = "048",
    year = "2023"
}

@article{Nordtvedt:1994gp,
    author = "Nordtvedt, K.",
    title = "{PostNewtonian gravity: Its theory - experiment interface}",
    doi = "10.1088/0264-9381/11/6A/009",
    journal = "Class. Quant. Grav.",
    volume = "11",
    pages = "A119--A132",
    year = "1994"
}

@article{Carlip:1997gf,
    author = "Carlip, Steven",
    title = "{Kinetic energy and the equivalence principle}",
    eprint = "gr-qc/9909014",
    archivePrefix = "arXiv",
    reportNumber = "UCD-97-10",
    doi = "10.1119/1.18885",
    journal = "Am. J. Phys.",
    volume = "66",
    pages = "409--413",
    year = "1998"
}

@article{Lebed:2011np,
    author = "Lebed, Andrei G.",
    title = "{Is Gravitational Mass of a Composite Quantum Body Equivalent to its Energy?}",
    eprint = "1311.2627",
    archivePrefix = "arXiv",
    primaryClass = "gr-qc",
    doi = "10.2478/s11534-013-0302-5",
    journal = "Central Eur. J. Phys.",
    volume = "11",
    pages = "969--976",
    year = "2013"
}

@article{Lebed:2014oua,
    author = "Lebed, Andrei G.",
    title = "{Breakdown of the equivalence between gravitational mass and energy for a composite quantum body}",
    eprint = "1404.4044",
    archivePrefix = "arXiv",
    primaryClass = "gr-qc",
    doi = "10.1088/1742-6596/490/1/012154",
    journal = "J. Phys. Conf. Ser.",
    volume = "490",
    pages = "012154",
    year = "2014"
}

@article{Zych:2018bmk,
    author = "Zych, Magdalena and Rudnicki, {\L}ukasz and Pikovski, Igor",
    title = "{Gravitational mass of composite systems}",
    eprint = "1808.05831",
    archivePrefix = "arXiv",
    primaryClass = "gr-qc",
    doi = "10.1103/PhysRevD.99.104029",
    journal = "Phys. Rev. D",
    volume = "99",
    number = "10",
    pages = "104029",
    year = "2019"
}

@article{Schwartz:2019qqg,
    author = "Schwartz, Philip K. and Giulini, Domenico",
    title = "{Post-Newtonian Hamiltonian description of an atom in a weak gravitational field}",
    eprint = "1908.06929",
    archivePrefix = "arXiv",
    primaryClass = "quant-ph",
    doi = "10.1103/PhysRevA.100.052116",
    journal = "Phys. Rev. A",
    volume = "100",
    number = "5",
    pages = "052116",
    year = "2019"
}

@article{Hartong:2023yxo,
    author = "Hartong, Jelle and Have, Emil and Obers, Niels A. and Pikovski, Igor",
    title = "{A coupling prescription for post-Newtonian corrections in quantum mechanics}",
    eprint = "2308.07373",
    archivePrefix = "arXiv",
    primaryClass = "gr-qc",
    doi = "10.21468/SciPostPhys.16.3.088",
    journal = "SciPost Phys.",
    volume = "16",
    number = "3",
    pages = "088",
    year = "2024"
}

@article{Janson:2024hdt,
    author = "Janson, Gregor and Lopp, Richard",
    title = "{General relativistic center-of-mass coordinates for composite quantum particles}",
    eprint = "2411.14307",
    archivePrefix = "arXiv",
    primaryClass = "gr-qc",
    doi = "10.1103/PhysRevD.111.064005",
    journal = "Phys. Rev. D",
    volume = "111",
    number = "6",
    pages = "064005",
    year = "2025"
}

@article{Sahoo:2018lxl,
    author = "Sahoo, Biswajit and Sen, Ashoke",
    title = "{Classical and Quantum Results on Logarithmic Terms in the Soft Theorem in Four Dimensions}",
    eprint = "1808.03288",
    archivePrefix = "arXiv",
    primaryClass = "hep-th",
    doi = "10.1007/JHEP02(2019)086",
    journal = "JHEP",
    volume = "02",
    pages = "086",
    year = "2019"
}

@article{Donoghue:1994dn,
    author = "Donoghue, John F.",
    title = "{General relativity as an effective field theory: The leading quantum corrections}",
    eprint = "gr-qc/9405057",
    archivePrefix = "arXiv",
    reportNumber = "UMHEP-408",
    doi = "10.1103/PhysRevD.50.3874",
    journal = "Phys. Rev. D",
    volume = "50",
    pages = "3874--3888",
    year = "1994"
}

@article{Bjerrum-Bohr:2014zsa,
    author = "Bjerrum-Bohr, N. E. J. and Donoghue, John F. and Holstein, Barry R. and Plant{\'e}, Ludovic and Vanhove, Pierre",
    title = "{Bending of Light in Quantum Gravity}",
    eprint = "1410.7590",
    archivePrefix = "arXiv",
    primaryClass = "hep-th",
    reportNumber = "IPHT-T14-108, IHES-P-14-33, ACFI-T14-21",
    doi = "10.1103/PhysRevLett.114.061301",
    journal = "Phys. Rev. Lett.",
    volume = "114",
    number = "6",
    pages = "061301",
    year = "2015"
}

@book{Weinberg:1995mt,
    author = "Weinberg, Steven",
    title = "{The Quantum theory of fields. Vol. 1: Foundations}",
    doi = "10.1017/CBO9781139644167",
    isbn = "978-0-521-67053-1, 978-0-511-25204-4",
    publisher = "Cambridge University Press",
    month = "6",
    year = "2005"
}

@article{Sundrum:2003jq,
    author = "Sundrum, Raman",
    title = "{Fat gravitons, the cosmological constant and submillimeter tests}",
    eprint = "hep-th/0306106",
    archivePrefix = "arXiv",
    doi = "10.1103/PhysRevD.69.044014",
    journal = "Phys. Rev. D",
    volume = "69",
    pages = "044014",
    year = "2004"
}

@article{Bern:2014oka,
    author = "Bern, Zvi and Davies, Scott and Nohle, Josh",
    title = "{On Loop Corrections to Subleading Soft Behavior of Gluons and Gravitons}",
    eprint = "1405.1015",
    archivePrefix = "arXiv",
    primaryClass = "hep-th",
    reportNumber = "UCLA-14-TEP-102",
    doi = "10.1103/PhysRevD.90.085015",
    journal = "Phys. Rev. D",
    volume = "90",
    number = "8",
    pages = "085015",
    year = "2014"
}

@article{Chisholm:1961tha,
    author = "Chisholm, J. S. R.",
    title = "{Change of variables in quantum field theories}",
    doi = "10.1016/0029-5582(61)90106-7",
    journal = "Nucl. Phys.",
    volume = "26",
    number = "3",
    pages = "469--479",
    year = "1961"
}

@article{Kamefuchi:1961sb,
    author = "Kamefuchi, S. and O'Raifeartaigh, L. and Salam, Abdus",
    title = "{Change of variables and equivalence theorems in quantum field theories}",
    doi = "10.1016/0029-5582(61)90056-6",
    journal = "Nucl. Phys.",
    volume = "28",
    pages = "529--549",
    year = "1961"
}

@article{Jacob:1959at,
    author = "Jacob, M. and Wick, G. C.",
    title = "{On the General Theory of Collisions for Particles with Spin}",
    doi = "10.1016/0003-4916(59)90051-X",
    journal = "Annals Phys.",
    volume = "7",
    pages = "404--428",
    year = "1959"
}

@article{Aoude:2023fdm,
    author = "Aoude, Rafael and Ochirov, Alexander",
    title = "{Gravitational partial-wave absorption from scattering amplitudes}",
    eprint = "2307.07504",
    archivePrefix = "arXiv",
    primaryClass = "hep-th",
    reportNumber = "IRMP-CP3-23-35",
    doi = "10.1007/JHEP12(2023)103",
    journal = "JHEP",
    volume = "12",
    pages = "103",
    year = "2023"
}

@article{Gasperini:1988zf,
    author = "Gasperini, M.",
    title = "{Testing the Principle of Equivalence with Neutrino Oscillations}",
    doi = "10.1103/PhysRevD.38.2635",
    journal = "Phys. Rev. D",
    volume = "38",
    pages = "2635--2637",
    year = "1988"
}

@article{Low:1958sn,
    author = "Low, F. E.",
    title = "{Bremsstrahlung of very low-energy quanta in elementary particle collisions}",
    doi = "10.1103/PhysRev.110.974",
    journal = "Phys. Rev.",
    volume = "110",
    pages = "974--977",
    year = "1958"
}

@article{Gross:1968in,
    author = "Gross, David J. and Jackiw, Roman",
    title = "{Low-Energy Theorem for Graviton Scattering}",
    doi = "10.1103/PhysRev.166.1287",
    journal = "Phys. Rev.",
    volume = "166",
    pages = "1287--1292",
    year = "1968"
}

@article{Elvang:2016qvq,
    author = "Elvang, Henriette and Jones, Callum R. T. and Naculich, Stephen G.",
    title = "{Soft Photon and Graviton Theorems in Effective Field Theory}",
    eprint = "1611.07534",
    archivePrefix = "arXiv",
    primaryClass = "hep-th",
    doi = "10.1103/PhysRevLett.118.231601",
    journal = "Phys. Rev. Lett.",
    volume = "118",
    number = "23",
    pages = "231601",
    year = "2017"
}

@article{Coleman:1967ad,
    author = "Coleman, Sidney R. and Mandula, J.",
    editor = "Zichichi, A.",
    title = "{All Possible Symmetries of the S Matrix}",
    doi = "10.1103/PhysRev.159.1251",
    journal = "Phys. Rev.",
    volume = "159",
    pages = "1251--1256",
    year = "1967"
}

@article{Tarallo:2014oaa,
    author = "Tarallo, M. G. and Mazzoni, T. and Poli, N. and Sutyrin, D. V. and Zhang, X. and Tino, G. M.",
    title = "{Test of Einstein Equivalence Principle for 0-spin and half-integer-spin atoms: Search for spin-gravity coupling effects}",
    eprint = "1403.1161",
    archivePrefix = "arXiv",
    primaryClass = "physics.atom-ph",
    doi = "10.1103/PhysRevLett.113.023005",
    journal = "Phys. Rev. Lett.",
    volume = "113",
    pages = "023005",
    year = "2014"
}

@article{Duan:2015zmf,
    author = "Duan, Xiao-Chun and Zhou, Min-Kang and Deng, Xiao-Bing and Yao, Hui-Bin and Shao, Cheng-Gang and Luo, Jun and Hu, Zhong-Kun",
    title = "{Test of the universality of free fall with atoms in different spin Orientations}",
    eprint = "1503.00433",
    archivePrefix = "arXiv",
    primaryClass = "physics.atom-ph",
    doi = "10.1103/PhysRevLett.117.023001",
    journal = "Phys. Rev. Lett.",
    volume = "117",
    pages = "023001",
    year = "2016"
}

@article{He:2014bga,
    author = "He, Song and Huang, Yu-tin and Wen, Congkao",
    title = "{Loop Corrections to Soft Theorems in Gauge Theories and Gravity}",
    eprint = "1405.1410",
    archivePrefix = "arXiv",
    primaryClass = "hep-th",
    reportNumber = "QMUL-PH-14-12",
    doi = "10.1007/JHEP12(2014)115",
    journal = "JHEP",
    volume = "12",
    pages = "115",
    year = "2014"
}

@article{Laddha:2018myi,
    author = "Laddha, Alok and Sen, Ashoke",
    title = "{Logarithmic Terms in the Soft Expansion in Four Dimensions}",
    eprint = "1804.09193",
    archivePrefix = "arXiv",
    primaryClass = "hep-th",
    doi = "10.1007/JHEP10(2018)056",
    journal = "JHEP",
    volume = "10",
    pages = "056",
    year = "2018"
}

@article{Krishna:2023fxg,
    author = "Krishna, Hare and Sahoo, Biswajit",
    title = "{Universality of loop corrected soft theorems in 4d}",
    eprint = "2308.16807",
    archivePrefix = "arXiv",
    primaryClass = "hep-th",
    reportNumber = "YITP-SB-2023-26",
    doi = "10.1007/JHEP11(2023)233",
    journal = "JHEP",
    volume = "11",
    pages = "233",
    year = "2023"
}

@article{Mirbabayi:2016xvc,
    author = "Mirbabayi, Mehrdad and Simonovi{\'c}, Marko",
    title = "{Weinberg Soft Theorems from Weinberg Adiabatic Modes}",
    eprint = "1602.05196",
    archivePrefix = "arXiv",
    primaryClass = "hep-th",
    month = "2",
    year = "2016"
}

@article{DeLuca:2024cjl,
    author = "De Luca, Valerio and Khoury, Justin and Wong, Sam S. C.",
    title = "{Gravitational memory and soft theorems: The local perspective}",
    eprint = "2412.01910",
    archivePrefix = "arXiv",
    primaryClass = "gr-qc",
    doi = "10.1103/gbg1-mz49",
    journal = "Phys. Rev. D",
    volume = "112",
    number = "2",
    pages = "L021502",
    year = "2025"
}

@book{Misner:1973prb,
    author = "Misner, Charles W. and Thorne, K. S. and Wheeler, J. A.",
    title = "{Gravitation}",
    isbn = "978-0-7167-0344-0, 978-0-691-17779-3",
    publisher = "W. H. Freeman",
    address = "San Francisco",
    year = "1973"
}

@article{Ni:1978zz,
    author = "Ni, Wei-Tou and Zimmermann, Mark",
    title = "{Inertial and gravitational effects in the proper reference frame of an accelerated, rotating observer}",
    doi = "10.1103/PhysRevD.17.1473",
    journal = "Phys. Rev. D",
    volume = "17",
    pages = "1473--1476",
    year = "1978"
}

@article{Newton:1949cq,
    author = "Newton, T. D. and Wigner, Eugene P.",
    title = "{Localized States for Elementary Systems}",
    doi = "10.1103/RevModPhys.21.400",
    journal = "Rev. Mod. Phys.",
    volume = "21",
    pages = "400--406",
    year = "1949"
}

@article{Foldy:1956kvk,
    author = "Foldy, Leslie L.",
    title = "{Synthesis of Covariant Particle Equations}",
    doi = "10.1103/PhysRev.102.568",
    journal = "Phys. Rev.",
    volume = "102",
    number = "2",
    pages = "568",
    year = "1956"
}

@article{Hehl:1990nf,
    author = "Hehl, Friedrich W. and Ni, Wei-Tou",
    title = "{Inertial effects of a Dirac particle}",
    doi = "10.1103/PhysRevD.42.2045",
    journal = "Phys. Rev. D",
    volume = "42",
    pages = "2045--2048",
    year = "1990"
}

@article{Zych:2011hu,
    author = "Zych, Magdalena and Costa, Fabio and Pikovski, Igor and Brukner, Caslav",
    title = "{Quantum interferometric visibility as a witness of general relativistic proper time}",
    eprint = "1105.4531",
    archivePrefix = "arXiv",
    primaryClass = "quant-ph",
    doi = "10.1038/ncomms1498",
    journal = "Nature Commun.",
    volume = "2",
    pages = "505",
    year = "2011"
}

@article{Pikovski:2013qwa,
    author = "Pikovski, Igor and Zych, Magdalena and Costa, Fabio and Brukner, Caslav",
    title = "{Universal decoherence due to gravitational time dilation}",
    eprint = "1311.1095",
    archivePrefix = "arXiv",
    primaryClass = "quant-ph",
    doi = "10.1038/nphys3366",
    journal = "Nature Phys.",
    volume = "11",
    pages = "668--672",
    year = "2015"
}

@article{Krajcik:1974nv,
    author = "Krajcik, R. A. and Foldy, L. L.",
    title = "{Relativistic center-of-mass variables for composite systems with arbitrary internal interactions}",
    doi = "10.1103/PhysRevD.10.1777",
    journal = "Phys. Rev. D",
    volume = "10",
    pages = "1777--1795",
    year = "1974"
}

@article{Sonnleitner:2018xbk,
    author = "Sonnleitner, Matthias and Barnett, Stephen M.",
    title = "{Mass-energy and anomalous friction in quantum optics}",
    eprint = "1806.00234",
    archivePrefix = "arXiv",
    primaryClass = "physics.atom-ph",
    doi = "10.1103/PhysRevA.98.042106",
    journal = "Phys. Rev. A",
    volume = "98",
    number = "4",
    pages = "042106",
    year = "2018"
}

@article{Mashhoon:1988zz,
    author = "Mashhoon, Bahram",
    title = "{Neutron interferometry in a rotating frame of reference}",
    doi = "10.1103/PhysRevLett.61.2639",
    journal = "Phys. Rev. Lett.",
    volume = "61",
    pages = "2639--2642",
    year = "1988"
}

@article{Parker:1980kw,
    author = "Parker, L.",
    title = "{ONE ELECTRON ATOM AS A PROBE OF SPACE-TIME CURVATURE}",
    doi = "10.1103/PhysRevD.22.1922",
    journal = "Phys. Rev. D",
    volume = "22",
    pages = "1922--1934",
    year = "1980"
}

@article{Goldberger:2004jt,
    author = "Goldberger, Walter D. and Rothstein, Ira Z.",
    title = "{An Effective field theory of gravity for extended objects}",
    eprint = "hep-th/0409156",
    archivePrefix = "arXiv",
    reportNumber = "UCSD-PTH-04-17, CMU-HEP-04-06",
    doi = "10.1103/PhysRevD.73.104029",
    journal = "Phys. Rev. D",
    volume = "73",
    pages = "104029",
    year = "2006"
}

@article{Giulini:2022wyp,
    author = "Giulini, Domenico and Gro{\ss}ardt, Andr{\'e} and Schwartz, Philip K.",
    title = "{Coupling Quantum Matter and Gravity}",
    eprint = "2207.05029",
    archivePrefix = "arXiv",
    primaryClass = "gr-qc",
    doi = "10.1007/978-3-031-31520-6_16",
    journal = "Lect. Notes Phys.",
    volume = "1017",
    pages = "491--550",
    year = "2023"
}

@article{PhysRevA.36.2763,
    author = "Lamb, Jr., Willis E. and Schlicher, Rainer R. and Scully, Marlan O.",
    title = "{Matter-field interaction in atomic physics and quantum optics}",
    journal = "Phys. Rev. A",
    volume = "36",
    number = "6",
    pages = "2763--2772",
    doi = "10.1103/PhysRevA.36.2763",
    year = "1987"
}

@article{DiStefano:2018fsc,
    author = "Di Stefano, Omar and Settineri, Alessio and Macr{\`\i}, Vincenzo and Garziano, Luigi and Stassi, Roberto and Savasta, Salvatore and Nori, Franco",
    title = "{Resolution of gauge ambiguities in ultrastrong-coupling cavity quantum electrodynamics}",
    eprint = "1809.08749",
    archivePrefix = "arXiv",
    primaryClass = "quant-ph",
    doi = "10.1038/s41567-019-0534-4",
    journal = "Nature Phys.",
    volume = "15",
    number = "8",
    pages = "803--808",
    year = "2019"
}

@article{Stokes:2019upq,
    author = "Stokes, Adam and Nazir, Ahsan",
    title = "{Gauge ambiguities imply Jaynes-Cummings physics remains valid in ultrastrong coupling QED}",
    eprint = "1805.06356",
    archivePrefix = "arXiv",
    primaryClass = "quant-ph",
    doi = "10.1038/s41467-018-08101-0",
    journal = "Nature Commun.",
    volume = "10",
    pages = "499",
    month = "1",
    year = "2019"
}

@article{Berry:1984jv,
    author = "Berry, Michael V.",
    title = "{Quantal phase factors accompanying adiabatic changes}",
    doi = "10.1098/rspa.1984.0023",
    journal = "Proc. Roy. Soc. Lond. A",
    volume = "392",
    pages = "45--57",
    year = "1984"
}

@article{Wilczek:1984dh,
    author = "Wilczek, Frank and Zee, A.",
    title = "{Appearance of Gauge Structure in Simple Dynamical Systems}",
    reportNumber = "NSF-ITP-84-51",
    doi = "10.1103/PhysRevLett.52.2111",
    journal = "Phys. Rev. Lett.",
    volume = "52",
    pages = "2111--2114",
    year = "1984"
}

@article{Kolodrubetz:2017ofs,
    author = "Kolodrubetz, Michael and Sels, Dries and Mehta, Pankaj and Polkovnikov, Anatoli",
    title = "{Geometry and non-adiabatic response in quantum and classical systems}",
    doi = "10.1016/j.physrep.2017.07.001",
    journal = "Phys. Rept.",
    volume = "697",
    pages = "1--87",
    year = "2017"
}

@article{Damour:2010rp,
    author = "Damour, Thibault and Donoghue, John F.",
    title = "{Equivalence Principle Violations and Couplings of a Light Dilaton}",
    eprint = "1007.2792",
    archivePrefix = "arXiv",
    primaryClass = "gr-qc",
    doi = "10.1103/PhysRevD.82.084033",
    journal = "Phys. Rev. D",
    volume = "82",
    pages = "084033",
    year = "2010"
}

@article{Armendariz-Picon:2011ydk,
    author = "Armendariz-Picon, Cristian and Penco, Riccardo",
    title = "{Quantum Equivalence Principle Violations in Scalar-Tensor Theories}",
    eprint = "1108.6028",
    archivePrefix = "arXiv",
    primaryClass = "hep-th",
    doi = "10.1103/PhysRevD.85.044052",
    journal = "Phys. Rev. D",
    volume = "85",
    pages = "044052",
    year = "2012"
}

@article{DiVecchia:2015jaq,
    author = "Di Vecchia, Paolo and Marotta, Raffaele and Mojaza, Matin and Nohle, Josh",
    title = "{New soft theorems for the gravity dilaton and the Nambu-Goldstone dilaton at subsubleading order}",
    eprint = "1512.03316",
    archivePrefix = "arXiv",
    primaryClass = "hep-th",
    reportNumber = "NORDITA-2015-134",
    doi = "10.1103/PhysRevD.93.085015",
    journal = "Phys. Rev. D",
    volume = "93",
    number = "8",
    pages = "085015",
    year = "2016"
}

@article{Hertzberg:2017nzl,
    author = "Hertzberg, Mark P. and Sandora, McCullen",
    title = "{Special Relativity from Soft Gravitons}",
    eprint = "1704.05071",
    archivePrefix = "arXiv",
    primaryClass = "hep-th",
    doi = "10.1103/PhysRevD.96.084048",
    journal = "Phys. Rev. D",
    volume = "96",
    number = "8",
    pages = "084048",
    year = "2017"
}

@article{Roura:2018cfg,
    author = "Roura, Albert",
    title = "{Gravitational redshift in quantum-clock interferometry}",
    eprint = "1810.06744",
    archivePrefix = "arXiv",
    primaryClass = "physics.atom-ph",
    doi = "10.1103/PhysRevX.10.021014",
    journal = "Phys. Rev. X",
    volume = "10",
    number = "2",
    pages = "021014",
    year = "2020"
}

@article{Cheinet:2005dk,
    author = "Cheinet, P. and Canuel, B. and Pereira Dos Santos, F. and Gauguet, A. and Yver-Leduc, F. and Landragin, A.",
    title = "{Measurement of the Sensitivity Function in a Time-Domain Atomic Interferometer}",
    eprint = "physics/0510197",
    archivePrefix = "arXiv",
    doi = "10.1109/TIM.2007.915148",
    journal = "IEEE Trans. Instrum. Meas.",
    volume = "57",
    number = "6",
    pages = "1141--1148",
    year = "2008"
}

@article{Giacomini:2020ahk,
    author = "Giacomini, Flaminia and Brukner, {\v{C}}aslav",
    title = "{Einstein's Equivalence principle for superpositions of gravitational fields and quantum reference frames}",
    eprint = "2012.13754",
    archivePrefix = "arXiv",
    primaryClass = "quant-ph",
    month = "12",
    year = "2020"
}

@inproceedings{Hardy:2019cef,
    author = "Hardy, Lucien",
    title = "{Implementation of the Quantum Equivalence Principle}",
    booktitle = "{Progress and Visions in Quantum Theory in View of Gravity}: {Bridging foundations of physics and mathematics}",
    eprint = "1903.01289",
    archivePrefix = "arXiv",
    primaryClass = "quant-ph",
    doi = "10.1007/978-3-030-38941-3_8",
    editor = "Finster, F. and Giulini, D. and Kleiner, J. and Tolksdorf, J.",
    pages = "189--220",
    publisher = "Birkh{\"a}user",
    address = "Cham",
    year = "2020"
}

@article{Draper:2020knh,
    author = "Draper, Tom and Knorr, Benjamin and Ripken, Chris and Saueressig, Frank",
    title = "{Graviton-Mediated Scattering Amplitudes from the Quantum Effective Action}",
    eprint = "2007.04396",
    archivePrefix = "arXiv",
    primaryClass = "hep-th",
    doi = "10.1007/JHEP11(2020)136",
    journal = "JHEP",
    volume = "11",
    pages = "136",
    year = "2020"
}

\end{document}